\documentclass[usegraphicx,useAMS,usenatbib]{mn2e}
\usepackage{ulem}
\usepackage{amssymb}

\title[Velocity Structure Diagnostics of Simulated Galaxy Clusters]{Velocity Structure Diagnostics of Simulated Galaxy Clusters}
\author[V. Biffi, K. Dolag and H. B\"ohringer]{V. Biffi$^{1,2}$\thanks{E--mail:
biffi@mpa-garching.mpg.de}, K. Dolag$^{1,3}$ and H. B\"ohringer$^2$\\
$^{1}$Max--Planck--Institut f\"ur Astrophysik, Karl--Schwarzschild--Strasse 1, D-85748 Garching bei M\"unchen, Germany\\
$^{2}$Max--Planck--Institut f\"ur extraterrestrische Physik, Giessenbachstrasse 1, D-85748 Garching bei M\"unchen, Germany\\
$^{3}$University Observatory Munich, Scheinerstr. 1, D-81679 M\"unchen, Germany}
\begin{document}
\def\openquote{``}
\def\closequote{''}
\def\tilde{~}
\def\fig{Fig.~}
\def\eq{Eq.~}
\def\sec{Section~}

\def\mfive{M_{500}}
\def\rfive{R_{500}}
\def\mtwom{M_{200m}}
\def\rtwom{R_{200m}}
\def\mtwo{M_{200}}
\def\rtwo{R_{200}}
\def\mtrue{M_{true}}
\def\mrot{M_{rot}}
\def\mstr{M_{str}}
\def\mth{M_{th}}

\def\msun{\rm{M_{\odot}}}
\def\kms{\rm{km/s}}
\def\mpc{\rm{Mpc}}
\def\kpc{\rm{kpc}}
\def\kev{\rm{keV}}

\def\lcdm{$\Lambda$CDM }
\def\om{\Omega_0}
\def\omb{\Omega_b}

\def\vtan{v_{tan}}

\def\mnras{MNRAS}
\def\apj{ApJ}
\def\apjs{ApJS}
\def\araa{ARA\&A}
\def\aa{A\&A}
\def\physrep{Phys. Rep.}

\date{Accepted .... Received ... ; ...}

\pagerange{\pageref{firstpage}--\pageref{lastpage}} \pubyear{...}

\maketitle

\label{firstpage}

\begin{abstract}
Gas motions in the hot intracluster medium of galaxy clusters have an
important effect on the mass determination of the clusters through
X--ray observations. The corresponding dynamical pressure has to be
accounted for in addition to the hydrostatic pressure support to
achieve a precise mass measurement. 
An analysis of the velocity structure of the ICM for simulated cluster--size 
haloes, especially focusing on rotational patterns, has been performed,
demonstrating them to be an intermittent phenomenon, strongly related to the 
internal dynamics of substructures. We find that 
the expected build--up of rotation due
to mass assembly gets easily destroyed by passages of gas--rich
substructures close to the central region. Though, if a typical
rotation pattern is established, the corresponding mass contribution
is estimated to be up to $\sim 17\%$ of the total mass in the innermost region, 
and one has to
account for it. Extending the analysis to a larger sample of simulated
haloes we statistically observe that (i) the distribution of the
rotational component of the gas velocity in the innermost region has
typical values of $\sim 200-300 \kms;$ (ii) except for few outliers,
there is no monotonic increase of the rotational velocity with
decreasing redshift, as we would expect from approaching a relaxed
configuration. Therefore, the hypothesis that the build--up of
rotation is strongly influenced by internal dynamics is confirmed, and
minor events like gas--rich substructures passing close to the
equatorial plane can easily destroy any ordered rotational pattern. 
\end{abstract}


\section{Introduction}
Within the hierarchical structure--formation scenario, 
galaxy clusters are key targets that
allow us to study both the dynamics on the gravity--dominated scale
and the complexity of astrophysical processes dominating on the small
scale. In such studies their mass is one of the most crucial
quantities to be evaluated, and the bulk properties measured from
X--ray observations still provide the best way to estimate the mass,
primarily on the assumption of hydrostatic equilibrium
\cite[][]{sarazin1988}. Mass estimates rely then on the assumptions
made about the cluster dynamical state, since the Hydrostatic
Equilibrium Hypothesis (HEH) implies that only the thermal pressure of
the hot ICM is taken into account \cite[][]{rasia04}. Lately, it has
been claimed in particular that non--thermal motions, as rotation,
could play a significant role in supporting the ICM in the innermost
region \cite[e.g. ][]{lau2009,fang2009} biasing the mass measurements
based on the HEH. The analysis of simulated cluster--like objects
provides a promising approach to get a better understanding of the
intrinsic structure of galaxy clusters and the role of gas dynamics,
which can be eventually compared to X--ray observations. Because of
the improvement of numerical simulations, the possibility to study in
detail the physics of clusters has enormously increased \cite[see][for 
a recent comprehensive review]{borgani2009} and future satellites
dedicated to high--precision X--ray spectroscopy, such as ASTRO--H and
IXO, will allow to detect these ordered motions of the ICM. With this
perspective, we perform a preliminary study on the ICM structure for
some clusters extracted from a large cosmological hydrodynamical
simulation, investigating in particular the presence of rotational
motion in the ICM velocity field.\\ 
\indent The paper is organized as follows. We describe the numerical
simulations from which the samples of cluster--like haloes have been selected 
in \sec\ref{secSims}. In \sec\ref{secVrot} we consider a first set of simulated 
clusters and present results on build--up of rotation in the halo core for single 
cases of study (\sec\ref{subsecRot} and \sec\ref{subsecg51}), while results about the 
contribution to the mass estimations are given in \sec\ref{secMrot}. A second sample 
of clusters is then statistically investigated in \sec\ref{secPadme}. We 
discuss our results and conclude in \sec\ref{secConclusion}.\\
\indent Appendix A is devoted to comment on the effects of artificial viscosity, 
while in Appendix B we briefely comment on the ellipticity profiles of the simulated clusters.
\section[]{Numerical Simulations} \label{secSims}
We consider two sets of cluster--like haloes selected from two
different parent cosmological boxes. In both cases the cosmological
simulations were performed with the TreePM/SPH code GADGET--2
\cite[][]{springel2001,springel2005}, assuming a slightly different
cosmological model (a standard \lcdm model and a WMAP3 cosmology,
respectively) but including the same physical processes governing the
gas component \cite[][]{springel2003}, i.e. radiative cooling, star formation, and supernova
feedback (\textit{csf} simulation, see \cite{dolag2009} and references
therein for a more detailed overview on different runs of the parent
hydrodynamical simulations we refer to in our work). Additionally, 
we refer to simulations of the same objects without including
radiative processes as \textit{ovisc}.\\ 
\indent
\textbf{Set 1.} The first data set considered consists of 9 cluster--size 
haloes, re--simulated with higher resolution using the \openquote zoomed 
initial condition\closequote (ZIC) technique \cite[][]{tormen1997}.
The clusters have been originally extracted from a
large--size cosmological simulation of a \lcdm universe with $\om =
0.3$, $h = 0.7$, $\sigma_8 = 0.9$ and $\omb = 0.039$, within a box of
$479 h^{-1} \mpc$ a side. The final mass--resolution of these simulations is 
$m_{DM}=1.13 \times 10^{9}h^{-1}\msun$ and $m_{gas}=1.69 \times 10^{8}h^{-1}\msun,$ 
for the DM and gas particles, respectively. The spatial resolution for Set 1 
reaches $5 h^{-1}\kpc$ in the central parts and for the most massive clusters we 
typically resolve up to 1000 self--bound sub--structures within $R_{vir}$ 
\cite[as shown in][]{dolag2009}.
The main haloes have masses larger than
$\sim1.1 \times 10^{14}h^{-1}\msun$ and they have all been selected in
a way that they are quite well--behaved spherically--shaped objects at
present epoch, although a fair range from isolated and potentially
relaxed objects to more disturbed systems embedded within larger
structures is available. 
Having a reasonable dense sample within the time domain (e.g. 50 outputs 
between $z=1$ and today) and high resolution, we can study how common and 
significant the rotational support of the ICM is. In particular we focus on 
the detailed evolution of such rotational motions. \\ 
\indent \textbf{Set 2.} 
In the second data set we analyzed a volume limited sample of cluster--size 
haloes, where we computed the distribution of the ICM rotational velocity 
and compared their distribution at different redshift. 
This second set of clusters has been extracted from a large
size cosmological simulation with a box--size of $300 h^{-1} \mpc,$ 
simulated with $2\times768^3$ particles, 
assuming the 3--year WMAP values for the cosmological parameters
\cite[][]{spergel2007}, i.e. $\om = 0.268$, $\omb = 0.044$, $\sigma_8
= 0.776$ and $h = 0.704$. The final mass--resolution for this second set of
simulations is $m_{DM}=3.71 \times 10^{9}h^{-1}\msun$ and 
$m_{gas}=7.28 \times 10^{8}h^{-1}\msun$. 
Given the larger sample, a fair
investigation of the amount of rotational support within the ICM 
from a statistical point of view is then possible.\\
\indent The two sets of simulations analyzed were performed with a different 
value of $\sigma_8,$ meaning that differences in the merging histories can be introduced. Nevertheless,
we stress that the purpose of the second set is only to enlarge the statistics on the build--up of 
rotational motions in simulated galaxy clusters, and not a direct comparison of single objects to
the objects of Set 1. Therefore, we are confident that our conclusions do not depend on these differences
in the parameters of the two simulations.

\section{VELOCITY STRUCTURE OF THE ICM} \label{secVrot}
\begin{figure*}
	\includegraphics[width=0.99\textwidth]{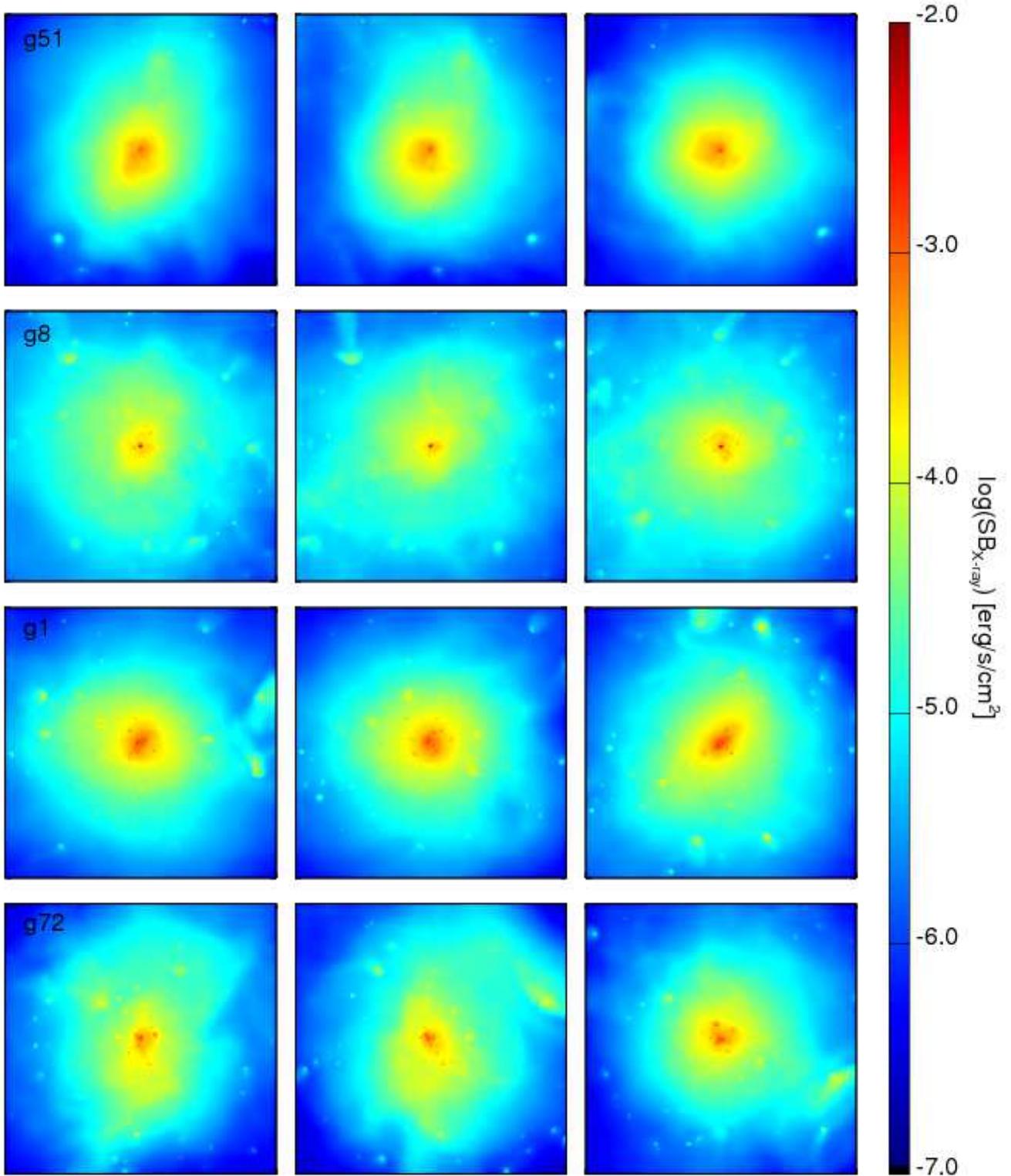}
	\caption{X--ray surface brightness maps along the three projection axis for the four 
most massive cluster--like haloes in Set 1 (from top to bottom: g51, g8, g1, g72). Each map is
$2\mpc$--side, enclosing therefore the region of about $\sim \rfive$.}
\label{hutt_maps}
\end{figure*}

\begin{figure}
	\includegraphics[scale=0.49]{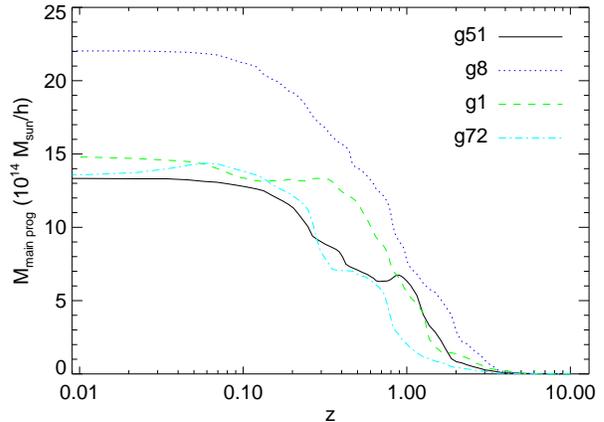}
	\caption{Main halo mass accretion history for the four most massive clusters in Set 1: 
g51 (\textit{solid, black line}), g8 (\textit{dotted, blue}), g1 (\textit{dashed, green}), 
and g72 (\textit{dot--dashed, light--blue}). 
The mass of the main halo is plotted in units of $10^{14}h^{-1}\msun$ as function of redshift. }
	\label{hutt_mainprog}
\end{figure}

In the first place, we have been investigating the velocity structure
of the ICM for all the main haloes of the Set 1, looking for evidence
of rotational patterns in the most central regions of the simulated
clusters, especially for the most massive haloes.\\
\indent Although a variety of objects is offered (ranging from more complex 
structures sitting in a denser environment to quite isolated haloes),
all the cluster-like haloes have been originally selected to be fairly regularly shaped.
For this aspect, the whole Set 1 is biased towards quite relaxed clusters and
we would expect to find a more significant amount of rotation than on average. 
Among the four most massive systems, for which \fig\ref{hutt_maps} gives the X--ray
surface brightness maps in the three projected directions,
we identify as the most relaxed clusters g51 and g8, while g1 and g72 are disturbed 
systems still suffering at the present epoch from recent major mergers 

\indent An indication of this differences is clearly shown in \fig\ref{hutt_mainprog}, where we plot
the mass accretion history for the main progenitor of each cluster as function of redshift.
The curves referring to g51 and g8 (solid black line and dotted blue line, respectively) 
show a smoother mass assembly (at late times, i.e. $z\lesssim 0.3$) 
if compared with those for g1 and g72 (dashed green line and 
dot--dashed light--blue line, respectively), whose curves show bumps related to major merging events 
down to very low redshift. 
In addition, a visual inspection of the X--ray surface brightness maps presented in \fig\ref{hutt_maps}
definitely suggests g51 to be the less sub-structured halo.\\
\indent In this perspective, we address g51 as the best
case of study to explore the build--up of rotational motions in the cluster central region 
as a consequence of the cooling of the core. Among the other clusters also g1 shows interesting
features in its velocity field that are worth to be investigated in more detail and compared to
the case of g51 to better characterize the occurrence of ordered rotational motions in
the intracluster gas
(see Appendix~\ref{appendix2} for a comment on the isophote ellipticities of g51 and g1
among the four most massive haloes from Set 1 
and how the gas shapes relate to intrinsic rotational gas motions).\\
\\
\subsection{Rotational patterns in the ICM}\label{subsecRot}
The two cases analyzed in detail (the isolated regular cluster, g51, and
the disturbed massive halo g1) are particularly interesting for our purpose,
since their velocity structure at redshift $z=0$ shows two opposite pictures, namely strong
rotational patterns for g1 and almost no gas rotation for g51.\\ 
\indent 
In the classic cooling flow model, gas rotation is expected near the center
of the flow because of mass and angular momentum conservation
\cite[e.g.][]{mathews2003}. Though, the rate of cooling gas predicted
by the classical paradigm of cooling flows is rarely observed in real
clusters, implying that feedback processes must play a role in
preventing cooling \cite[][]{macnamara2007}. In contrast, in
simulations, the strong cooling in the central region of the simulated
cluster--like haloes suggests that relaxed objects should build up
significant rotational motion in the innermost region where gas is
infalling and contracting under the conservation of angular
momentum. As reported in the literature \cite[e.g.][]{fang2009} this
effect is expected to be particularly evident in simulated clusters
that can be identified as relaxed objects. Therefore it is 
interesting to point out that, in our sample, not even in the object
with the smoothest accretion history and less substructured morphology 
significant rotational patterns establish in the ICM velocity structure as a
consequence of collapse.\\ 
\indent 
Dealing with hydrodynamical
simulations though, the build--up of rotation in the central
region of clusters can be also enhanced by an 
excess of gas cooling that has been found to overproduce the observed cosmic
abundance of stellar material
\cite[e.g][]{katzwhite1993,balogh2001} in absence of very strong, 
not yet fully understood feedback processes. In our simulations, the
implementation of a multi--phase model for star formation
\cite[e.g.][]{katz1996,springel2003} and the treatment of the thermal
feedback process, including also galactic winds associated to star
formation, is able to partially reduce the over--cooling problem \cite[][]{borgani2006}. This
fact plausibly contributes to prevent significant rotation.\\ 
\indent 
In order to study in detail the rotational component of the ICM
velocity for a halo, we first define a \openquote best equatorial
plane\closequote on which we can calculate the tangential component of the
velocity, $\vtan.$ This plane is taken to be perpendicular to the
direction of the mean gas angular momentum, $\textbf{j}$, calculated
averaging over the gas particles within the region where we want to
investigate the 
rotational motion within the ICM, i.e. $<0.1\rfive$ (here, the
overdensity of $500$ is defined with respect to the critical density
of the Universe). This definition of the \openquote best equatorial
plane\closequote allows us to emphasize and characterize the rotation
of the gas in an objective way for all clusters, whenever it 
appears.\\ 
\indent 
To perform our analysis, we rotate the halo such that
the new $z$--axis is aligned with the direction of $\textbf{j}$ and
the new $xy$ plane easily defines the best equatorial plane. After
subtracting an average bulk velocity for the gas component within the
region corresponding to $\rfive$, we compute the tangential component
of the velocity on this plane. We consider a $40h^{-1}\kpc$ slice of
the simulation box containing this plane for all the calculations
hereafter.

\subsection{A case study: g51 vs. g1} \label{subsecg51}
\begin{figure*}
\includegraphics[width=0.495\textwidth,height=0.3\textheight]{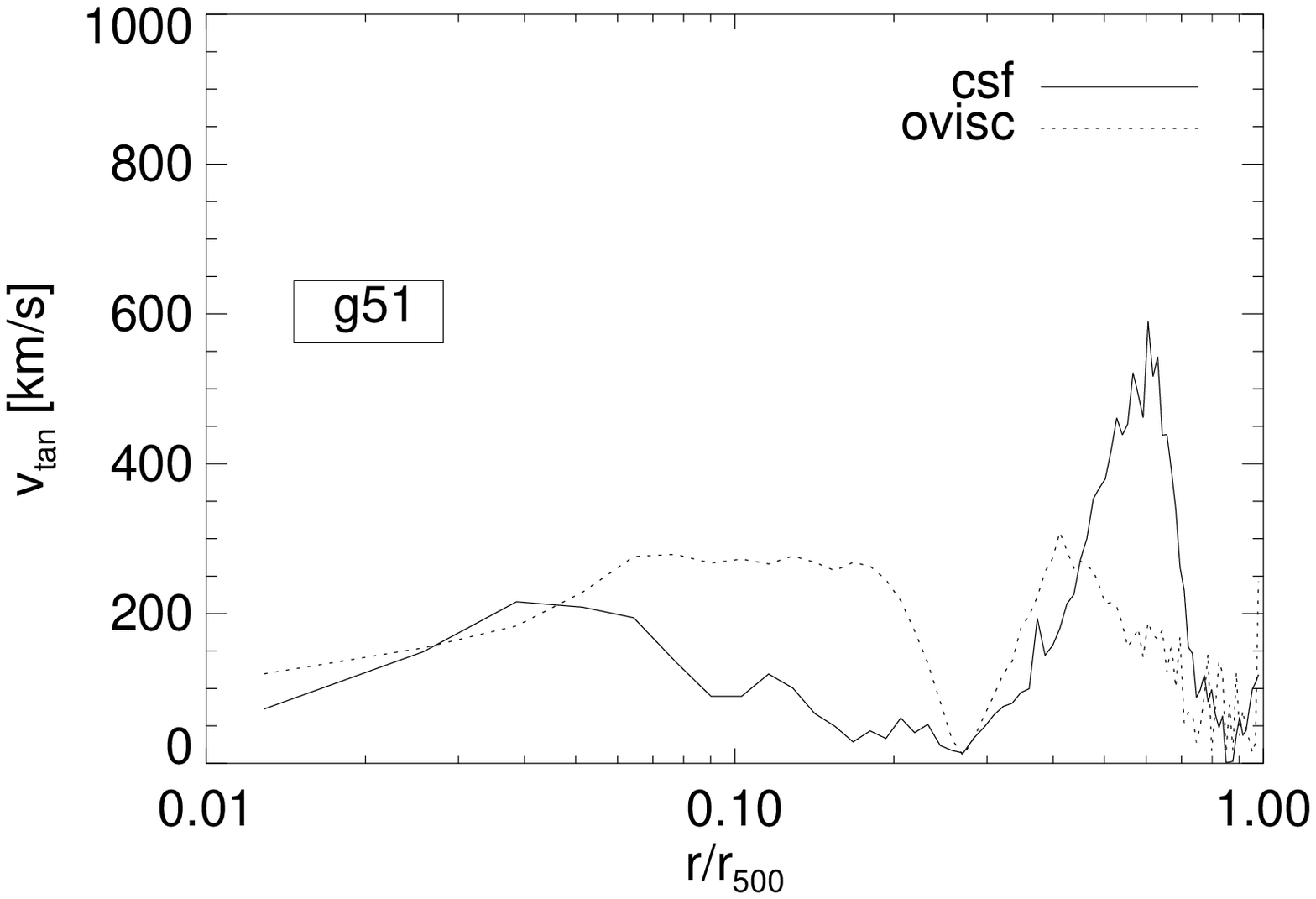}
\includegraphics[width=0.495\textwidth,height=0.3\textheight]{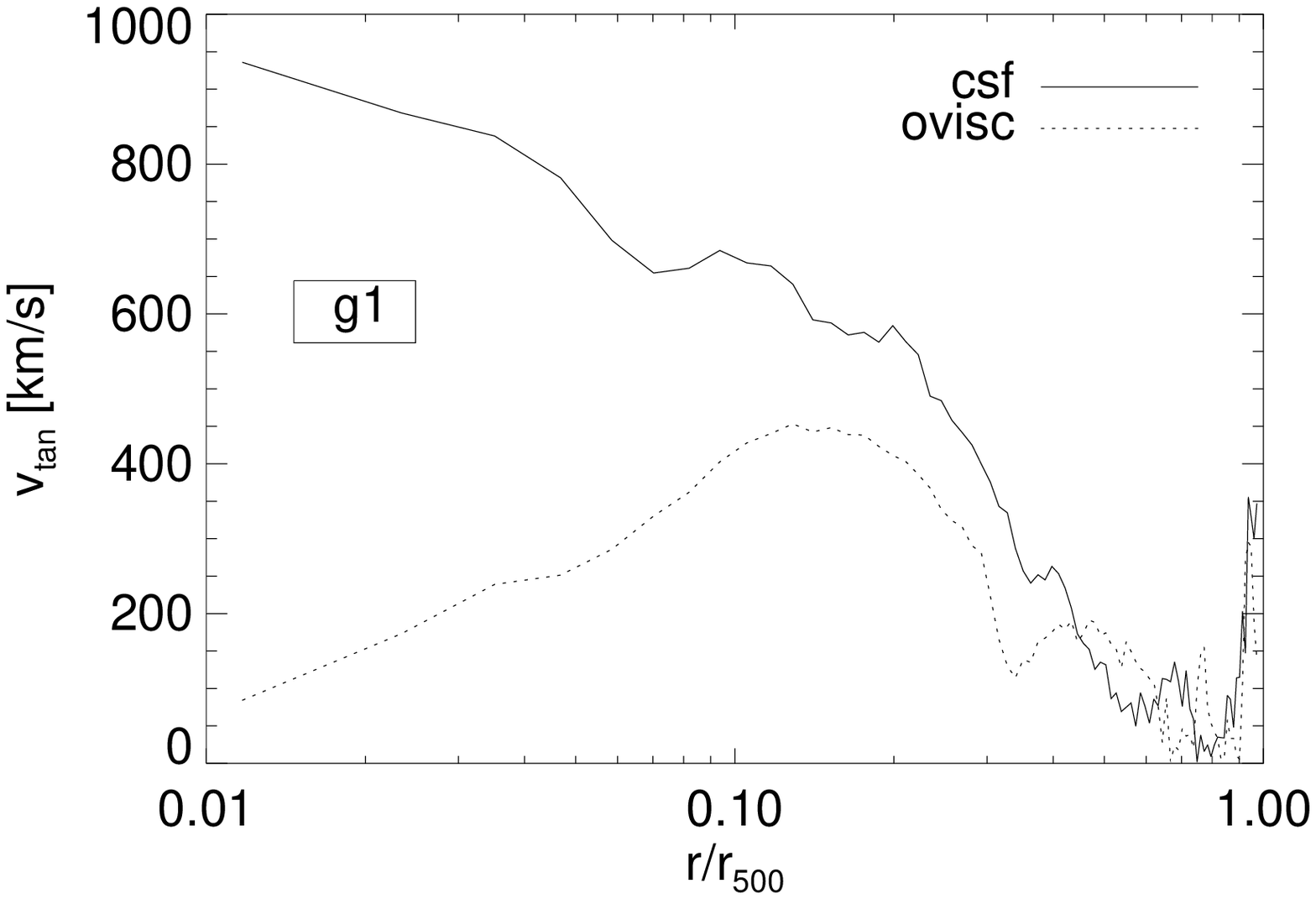}
\caption{Rotational velocity as function of the radius out to $\rfive$ for a relaxed cluster (g51, \textit{left panel}) and for a highly disturbed system (g1, \textit{right panel}), at $z=0$. The tangential component of the ICM velocity is calculated on the best equatorial plane, i.e. the plane perpendicular to the direction of the mean gas angular momentum in the region within $0.1\rfive$. Two runs are compared: the \textit{csf} simulation (\textit{solid line}), including radiative cooling, star formation and supernova feedback, and the \textit{ovisc} simulation (\textit{dashed line}), in which all these physical processes are omitted and only a treatment for artificial viscosity is considered.}
\label{fig_prof}
\end{figure*}

As a case study, we particularly focus on g51, an isolated massive
cluster with gravitational mass of $\mtwom=1.34 \times
10^{15}h^{-1}\msun$ and a size of $\rtwom=2.28 h^{-1}\mpc$ at $z=0$,
and we compare it with the other extreme case mentioned, g1, which is
instead a strongly disturbed system with $\mtwom=1.49 \times
10^{15}h^{-1}\msun$ and $\rtwom=2.36 h^{-1}\mpc.$ Here, $\rtwom$ is
defined as the radius enclosing the region with density equal to $200$
times the mean density of the Universe, and $\mtwom$ is the mass
within $\rtwom.$ $\mtwom$ has been used throughout our study as
reference quantity to select haloes, but we always carry out our
calculations by referring to $\rfive$ and $\mfive,$ where the
overdensity of $500$ is instead defined with respect to the critical
density of Universe, motivated by a possible comparison to real X--ray
observations. In the case of g51 and g1, we have $\rfive=1.09,1.20
h^{-1}\mpc$ and
$\mfive=7.46\times10^{14},9.98\times10^{14}h^{-1}\msun$ respectively,
at $z=0.$\\
\indent From the considerations made in \sec\ref{secVrot} about its shape and accretion
history, g51 is likely to be, in a global sense, the most relaxed object in the sample.
In spite of this, at $z=0$ the velocity structure of the ICM in the innermost region is far
from showing a clear rotational pattern as expected from a nearly
homogeneous collapse process. However it shows some rotational pattern
at intermediate redshift.\\
\indent 
In \fig\ref{fig_prof} we plot the rotation velocity profile, $\vtan
(r),$ for the two interesting cases (at $z=0$) out to $\rfive.$ As explained in
the previous Section, $\vtan$ is the tangential component of the ICM
velocity, calculated in the best equatorial plane. In order to compute
the radial profile displayed in \fig\ref{fig_prof}, we make use of
radial bins in the plane to calculate the mass--weighted average value
of $\vtan$ of the gas particles at each $r.$ We have chosen $14
h^{-1}\kpc,$ as optimal bin width on the base of both resolution and
statistical motivation.\\ 
\indent 
The radial profiles for the rotational component of the gas velocity 
reflect the presence of a non--negligible rotational pattern in the ICM of the
disturbed system, while no significant rotation is built up in the
relaxed one. The profile of g51 (left panel in the Figure, solid
curve) shows relatively low values at small radii, and increases
significantly only at radii larger than $\sim0.3\rfive$, where the
rotational component of the velocity is likely to be dominated by some
bulk rotational motions, plausibly related to a subhalo
orbiting in the main halo close to $\rfive$. 
The value of $\vtan$ decreases instead with
increasing $r$ for g1 (right panel in the Figure, solid curve), where
the rotational velocity reaches almost $1000 \kms$ in the innermost
region. Also, it is interesting to compare with the rotational
velocity profile for two counterpart haloes, simulated without
including star formation and cooling (dashed curves). In such
simulation, referred to as \textit{ovisc} simulation \cite[see][as an
  overview]{dolag2009} the overcooling problem is completely avoided
because no stars are formed at all, and no significant rotation is expected to
build up in the center of the cluster--like haloes. Let us note that
for g51, the curves referring to the two simulations have a
significantly similar trend, while for g1 the \textit{csf} simulation
(solid curve) and the non--radiative one diverge towards the center,
increasing in the former and decreasing in the latter. While major events
occurring close to $z=0$ in the merging history of g1 could explain
the high values found for $\vtan$ in the innermost region, no major
mergers happen to characterize the history of g51 at late
time. Therefore a further zoom onto g51 is required in order to
understand the details of the 
processes that lead to the
build--up or to the disruption of gas rotation in the halo core.

\begin{figure}
\includegraphics[scale=0.48]{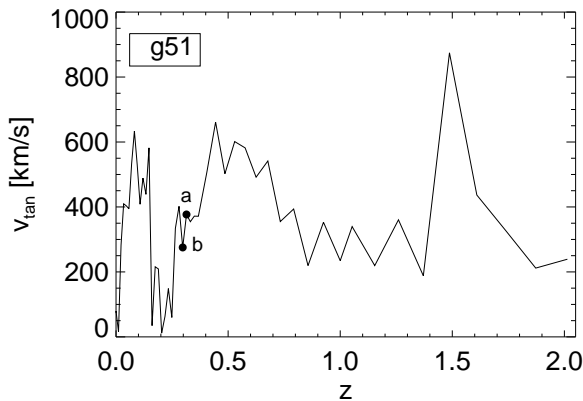}
\caption{ Evolution with redshift of the tangential component of the ICM velocity in the innermost region ($<0.1 \rfive$) of g51. }
\label{fig_evol}
\end{figure}

\begin{figure}
\includegraphics[scale=0.495]{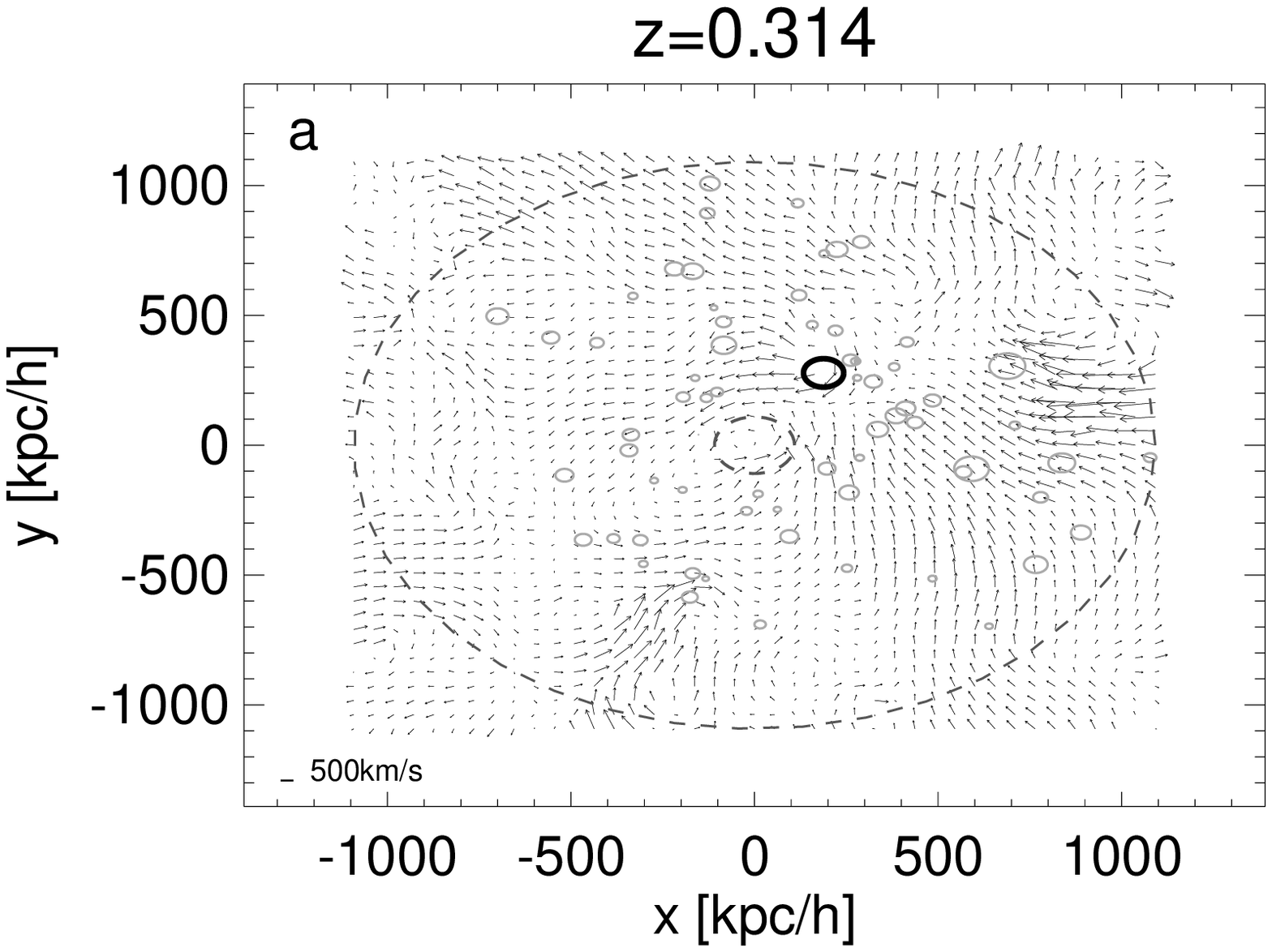}
\includegraphics[scale=0.495]{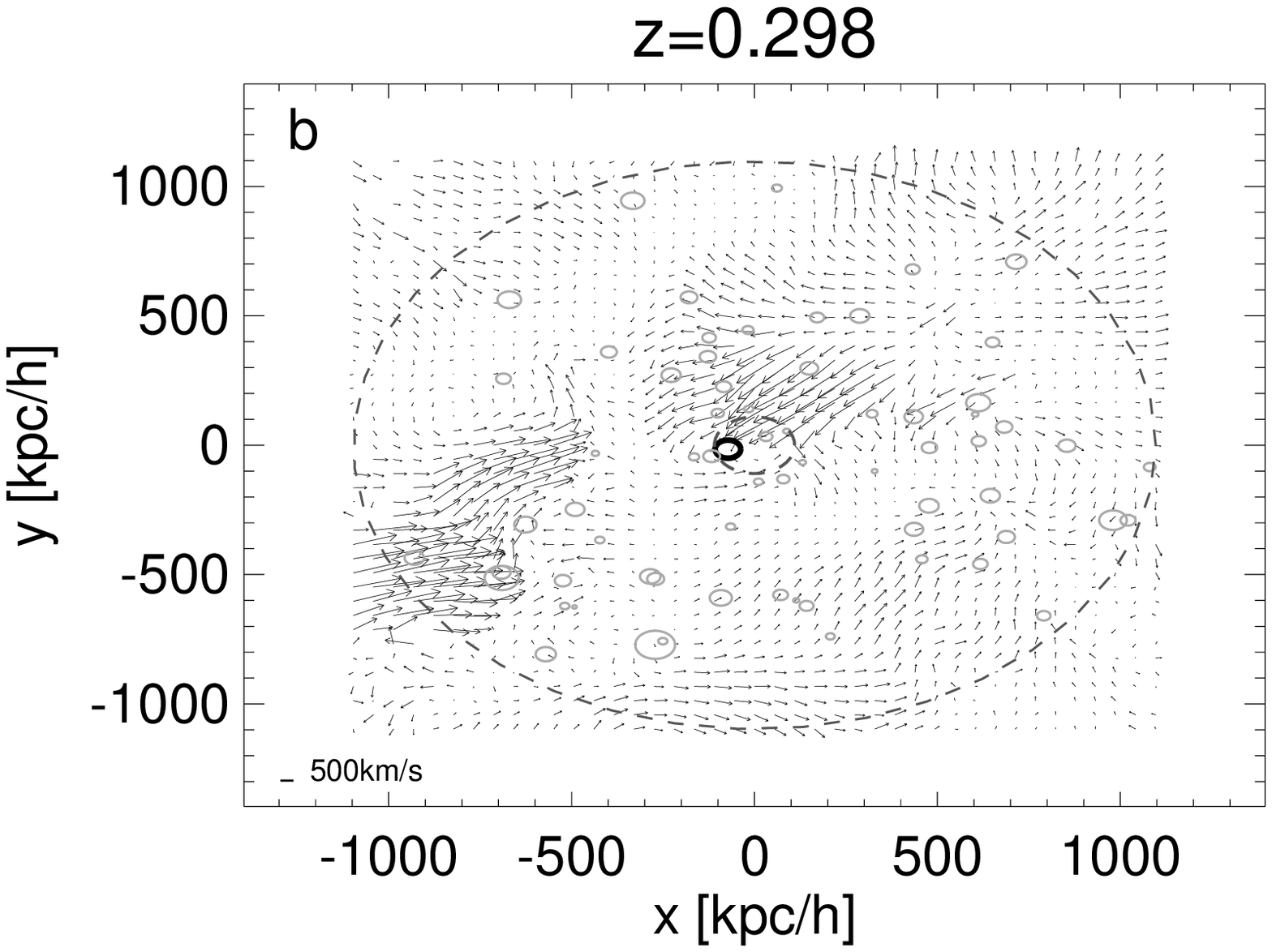}
\caption{Gas velocity fields at $z\sim 0.314$ (\textit{upper panel}) and $z\sim 0.298$ (\textit{lower panel}) projected onto the plane perpendicular to the direction of the gas mean angular momentum in the innermost region. The smaller and larger dashed circles mark respectively the regions of $0.1\rfive$ and $\rfive$, while the grey ones are DM--only subhaloes and the black circle is the gas--rich halo passing through the equatorial plane. The coordinates in the graphs are in comoving units.}
\label{fig_maps}
\end{figure}

\textbf{Rotational velocity evolution.} The possibility to track
back the history of the cluster--size haloes given by simulated data,
allows us to follow the redshift evolution of the rotational component
of the ICM velocity in the innermost region of g51, taken to be
$0.1\rfive$. Up to $z=2$, a mass--averaged value of the tangential
component of the ICM velocity has been calculated in the best
equatorial plane, so that rotation can be emphasized best whenever
there is one. 
At each redshift, the orientation of the best equatorial plane has been 
adjusted to be perpendicular to the direction of the 
mean gas angular momentum, as previously defined.\\
\indent 
While in the literature we find an inspiring work \cite[e.g.][]{fang2009} 
where values for the rotational velocity in
the central region of a relaxed cluster--like halo rise above
$1000\kms$, in our study this never happens and values generally
increase up to $650\kms$ as a maximum, except for high peaks probably
related to major merging events. 
These differences are likely to be related to the different amount of 
baryon cooling that characterizes the simulations analyzed in the work by
\cite{fang2009} \cite[extensively described in][]{kravtsov2005,kravtsov2006,nagai2007} 
with respect to those discussed here. The stellar fraction $f_{*}(<r) = M_{*}(<r)/M(<r)$
in the central part of our Set 1 clusters (i.e. $<\rfive$) is estimated to be
smaller than in \cite{fang2009} simulations, 
by about a factor of $\sim1.5$. 
Though, it is definitely higher than expected 
from observations of real clusters \cite[e.g.][]{lin2003}. 
The implementation of cooling in GADGET--2 reduces the overcooling problem,
naturally preventing strong rotation to get established.\\
\indent In \fig\ref{fig_evol} we plot the
variation of $\vtan$ with redshift, calculated in the innermost region
of g51. The peak shown around redshift $\sim 1.5$ is likely to be
driven by the last major merger occurring to g51, and is not related
to a quiescent build--up of mass and therefore of
rotation. Instead, within the redshift range $\sim 1.5-0.5$, a
general, although not smooth, increasing trend of $\vtan$ can be seen
in the plot, which is likely to be explained as the result of the
collapsing process under angular momentum conservation, although it is
difficult to show it quantitatively. At lower redshift, it is worth to
point out an interesting feature, that is the sudden drop of $\vtan$,
steeply decreasing twice at $z\sim 0$ and $\sim 0.3$. The breaks in
this expected general trend are not directly related to any major
event, and a deeper investigation of the ICM internal dynamics has
then been performed in order to understand the possible origin of this
unexpected behavior.
\\ \textbf{ICM velocity maps.} The panels in \fig\ref{fig_maps} show
the two--dimensional velocity field in the best equatorial plane in
the central slices of g51. Each velocity vector has a length
proportional to the absolute value of the velocity in that point of
the plane. The dashed circles mark the innermost region enclosed
within $0.1\rfive$ (smaller circle) and $\rfive$ (larger
circle).\\
\indent 
The velocity maps catch one of the two major decreases in the curve of
$\vtan$, in particular the one at roughly $z\sim 0.3$, which is the
first significant break in the increasing trend shown up to redshift
$\sim 0.5.$ Clearly, one can see the passage of a gas--rich subhalo
(thicker circle) through the best equatorial plane, onto which the gas
velocity field has been projected in the Figure. The subhalo is the
only gas-rich subhalo approaching the central region of the simulated
cluster.\\ 
\indent 
The two panels in \fig\ref{fig_maps} refer to redshift $z\sim 0.314$
(top) and $z\sim 0.297$ (bottom), and show the best moment right
before and after the first passage of the substructure through the
equatorial plane. The steep decrease of $\vtan$ does not start at this
moment nor does it reach the lowest value, but these two redshift
snapshots have been judged to best show a plausible explanation of the
suppression of rotation while it is happening. In fact, from the
velocity fields we notice that the gas shows a rotational motion with
velocities of $\sim380\kms$ in the innermost region, close to the
smaller dashed circle, while the subhalo is approaching (upper
panel). This rotational pattern is evidently disturbed in the lower
panel, where the subhalo has already passed through the plane, its gas
particles get probably stripped by the main halo gas and contribute to
decrease the velocity values to $\sim275\kms.$\\ 
\indent 
Let us stress that there are several DM--only substructures
permanently moving within the cluster and close to the innermost
region, but they do not disturb the build--up of rotation as
gas--rich subhaloes do.\\
\indent The decrease 
of $\vtan$ at redshift $\sim 0$ shows an analogous behavior.
\section{ROTATIONAL CONTRIBUTION TO TOTAL MASS} \label{secMrot}
\begin{figure*}
\includegraphics[scale=0.495]{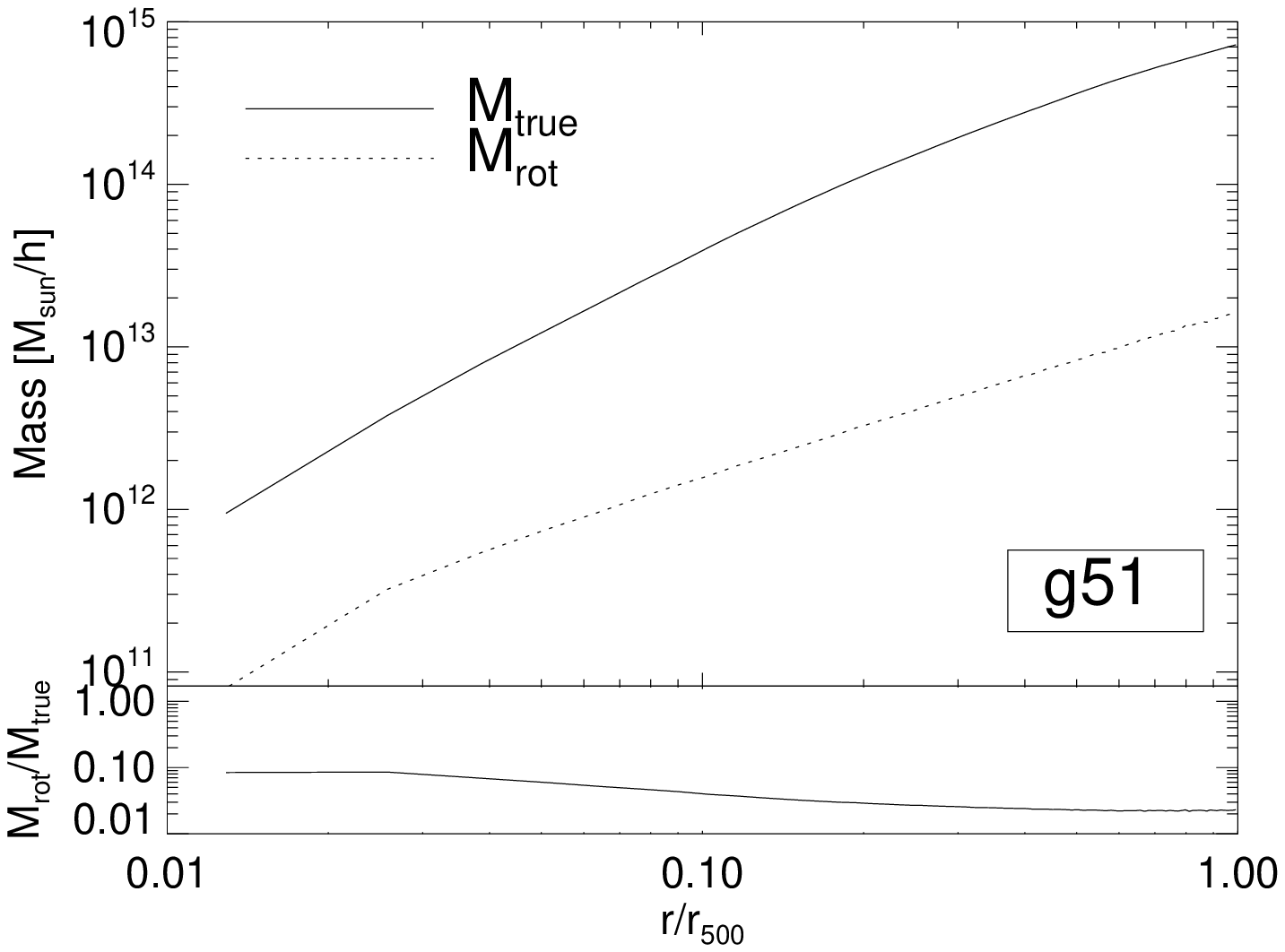}
\includegraphics[scale=0.495]{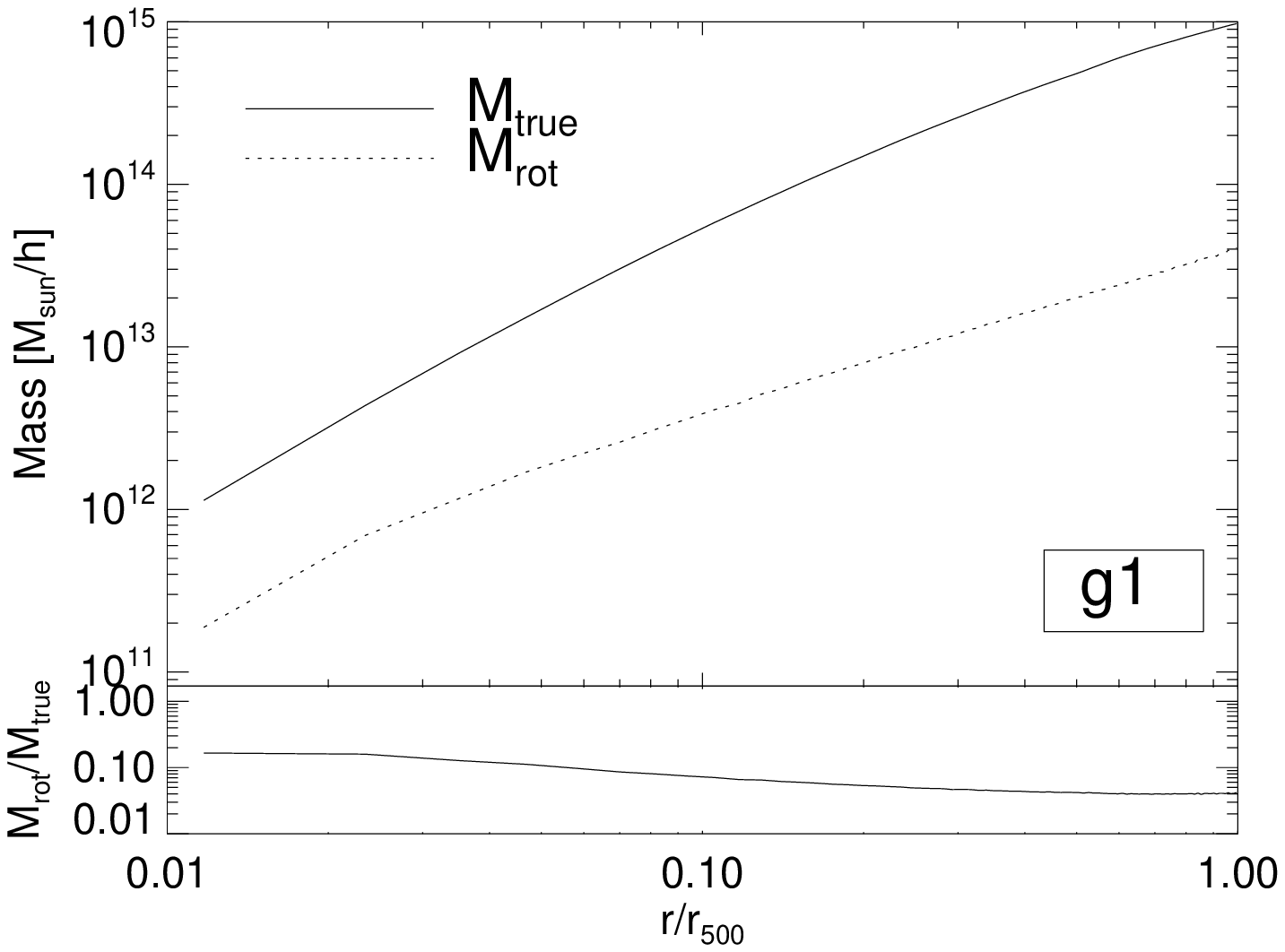}
\caption{Mass profiles of the relaxed cluster g51 (\textit{upper left panel}) and of the disturbed system g1 (\textit{upper right panel}), at $z=0$. Top: in each panel, the radial profiles of the true mass, $\mtrue$ (\textit{solid line}), and of the component estimated from the gas rotation, $\mrot$ (\textit{dotted line}), out to $\rfive$ are shown. Bottom: ratio of the two mass terms $\mrot/\mtrue$ as function of radius for g51 (\textit{lower left panel}) and g1(\textit{lower right panel}). }
\label{figmass}
\end{figure*}
In this section we compare 
the contribution coming from rotational motions that should be considered in the estimation of total mass 
with the total mass calculated for the simulated clusters.
Formally, the total cluster
mass $M$  enclosed within a closed surface $\textbf{S}$ is given by
Gauss's Law
\begin{equation}\label{gauss}
 M=\frac{1}{4\pi G} \int \nabla\Phi \cdot  d^2 \textbf{S},
\end{equation}
where $\Phi$ is the cluster gravitational potential and $G$ is the
gravitational constant. Under the assumptions that the
ICM is a steady--state, inviscid, collisional fluid, we can replace
the term $\nabla \Phi$ with the terms involving gas pressure and
velocity using Euler's equation as follows:
\begin{equation}\label{gauss_euler}
 M=\frac{1}{4\pi G} \int \left[ -\frac{1}{\rho_g}\nabla P_g -
   (\textbf{v} \cdot \nabla)\textbf{v} \right] \cdot  d^2 \textbf{S},
\end{equation}
where $\rho_g$ and $P_g$ are the gas density and pressure
respectively. While the pressure term within the integral represents
the contribution of the gas random motions, both thermal and
turbulent, the velocity term includes the ordered motions in the ICM,
i.e. rotational and streaming motions. For the purpose of our work, we
are mainly interested in the contribution to the pressure support
given by the rotational motions of the hot intracluster gas, and we
therefore separate the velocity term in \eq\ref{gauss_euler} into
\begin{equation}\label{mrot}
 \mrot=\frac{1}{4\pi G} \int \left( \frac{v_{\theta}^2 +v_{\phi}^2}{r}
 \right) d^2 S
\end{equation}
and
\begin{equation}\label{mstream}
 M_{str}=-\frac{1}{4\pi G} \int \left( v_r\frac{\partial v_r}{\partial
   r} + \frac{v_{\theta}}{r}\frac{\partial v_r}{\partial \theta} +
 \frac{v_{\phi}}{r \sin \theta}\frac{\partial v_r}{\partial \phi}
 \right) d^2 S,
\end{equation}
by evaluating $(\textbf{v} \cdot \nabla)\textbf{v}$ in spherical
coordinates \cite[][]{bt08,fang2009}. In particular, the term due to streaming
motion, $\mstr,$ is likely to be less relevant than $\mrot,$
especially for relaxed clusters. Therefore we explicitly calculate the
rotation term for our clusters g51 and g1 at several redshifts. In our
analysis we compare this term with the total \openquote
true\closequote mass of gas and dark matter for the cluster, $\mtrue,$
computed directly by summing up all the particle masses.\\ 
\indent 
In \fig\ref{figmass} we show the radial profiles of both $\mrot$ and
$\mtrue,$ and their ratio, for g51 (left panel) and g1 (right panel)
at $z=0.$ In order to evaluate \eq\ref{mrot} we consider the full
three--dimensional structure of the velocity field without any assumption
of spherical symmetry, as stated in Gauss's theorem, so that our
calculation is completely independent of the cluster geometry. In our
approximation of the integral that appears in \eq\ref{mrot}, the
closed surface $\textbf{S}$ has been replaced with a spherical shell
at each radial bin, and we sum over the single particle contributions
in the shell instead of dividing the surface in cells. Specifically,
at each radial bin we associate an effective area to each gas particle
in the shell and compute the integrand term considering the velocity
$(v_{\theta}^2 +v_{\phi}^2)$ of the particle. We make use of a radial
binning up to $\rfive,$ each bin of $14 h^{-1}\kpc,$ consistent with
the profiles of $\vtan (r)$ discussed in \sec\ref{subsecg51}.\\ \indent
The profiles shown in \fig\ref{figmass} have similar trends for both
g51 and g1, but one can clearly notice that the more significant
rotational pattern found in the innermost region of the disturbed
cluster g1 with respect to g51 is reflected here in a more significant
contribution to the total mass of $\mrot.$ In general, the rotation
appears to be more important in the innermost regions than in the
cluster outskirts. Indeed, out to $\rfive$ the rotational component of
the total mass accounts for $\sim2\%$ of the true mass, $\mtrue,$ in
g51 and slightly more in the case of g1 ($4\%$). In contrast, $\mrot$
plays a more significant role at radii $<0.1\rfive,$ where its
contribution can be up to $9-17\%$ of $\mtrue,$ the highest value
reached in the core region of g1. Although we consider purely
rotational motions, other non--thermal motions should be accounted
for as well and we can conclude that, if rotation establishes, it can
significantly contribute to the total pressure support to the cluster
weight.

\section{EXTENDING THE STATISTICS TO A LARGER SIMULATED SAMPLE} \label{secPadme}
\begin{figure*}
\begin{center}
\includegraphics[scale=0.6]{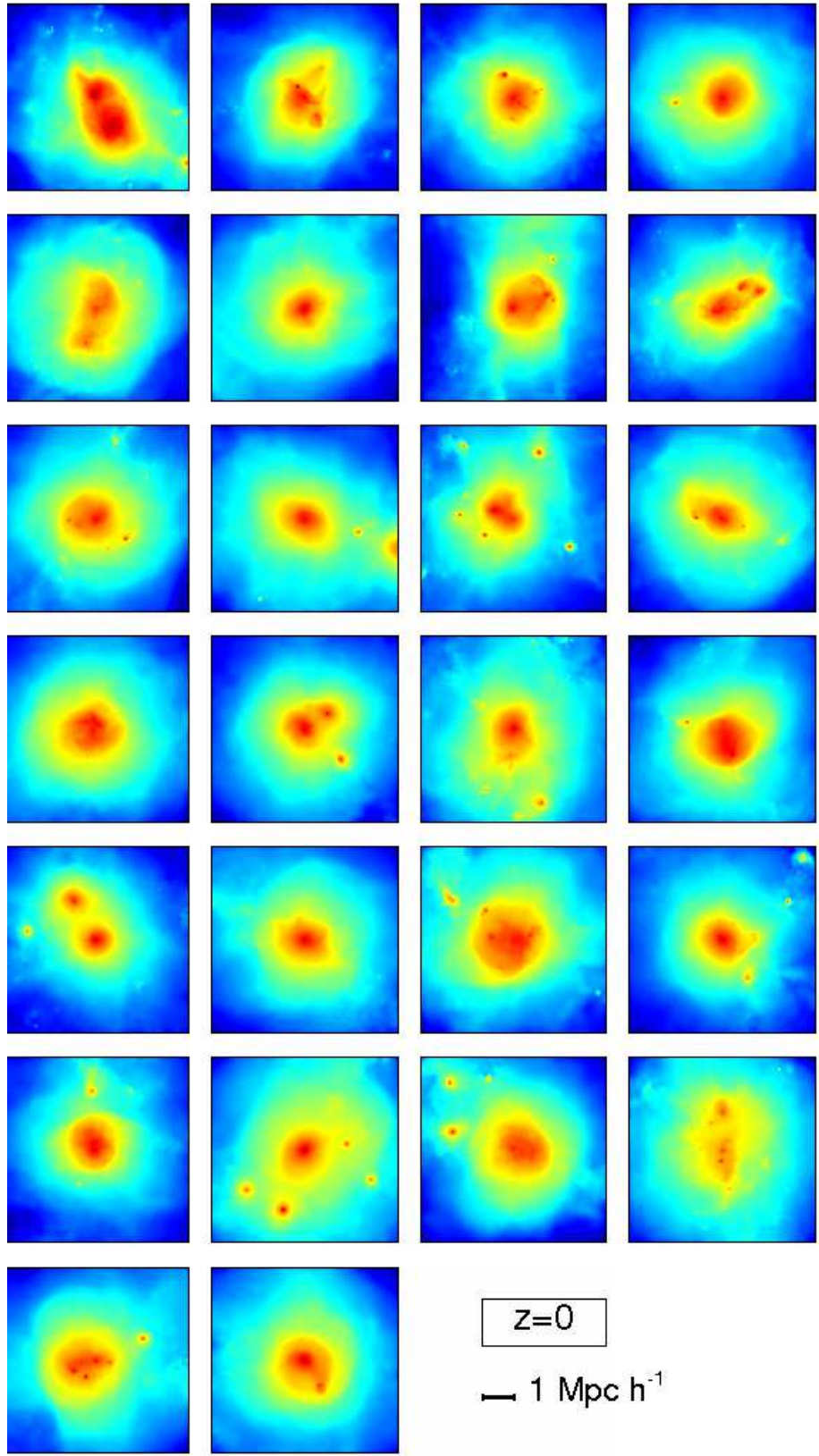}
\caption{Visualization of the gas SB, in arbitrary units, for the 26 cluster--size haloes selected at $z=0$ from the Set 2.}
\label{figpadme}
\end{center}
\end{figure*}
In order to gain a more reliable overview of the phenomenon of the
appearance of rotational patterns in the innermost regions of
cluster--like haloes, our analysis has also been performed for Set 2,
a larger sample of simulated objects spanning a range of $\mtwom$ from
$\sim 5 \times 10^{14}h^{-1}\msun$ to $\sim 2.2\times
10^{15}h^{-1}\msun.$\\ 
\indent 
The selection of the sample has been carried out such that we isolate
all the haloes in the cosmological box with virial mass above a mass
threshold, chosen to keep a statistically reliable number of objects
throughout the redshift range explored. At  $z=0,$ we selected 26
haloes with $\mtwom > 5 \times 10^{14}h^{-1}\msun,$ 
for which a visual
representation is given in \fig\ref{figpadme}. 
At higher redshift, the mass
threshold is lower in order to have a fair sample to investigate
statistically. We calculate the distribution of the
rotational velocity on the best equatorial plane in the innermost
region (i.e. $<0.1\rfive$) of each selected halo, sampling the
redshift range $[0,0.5]$.

\subsection{Distribution of rotational velocities at various redshifts}
\begin{figure*}
\includegraphics[scale=0.49]{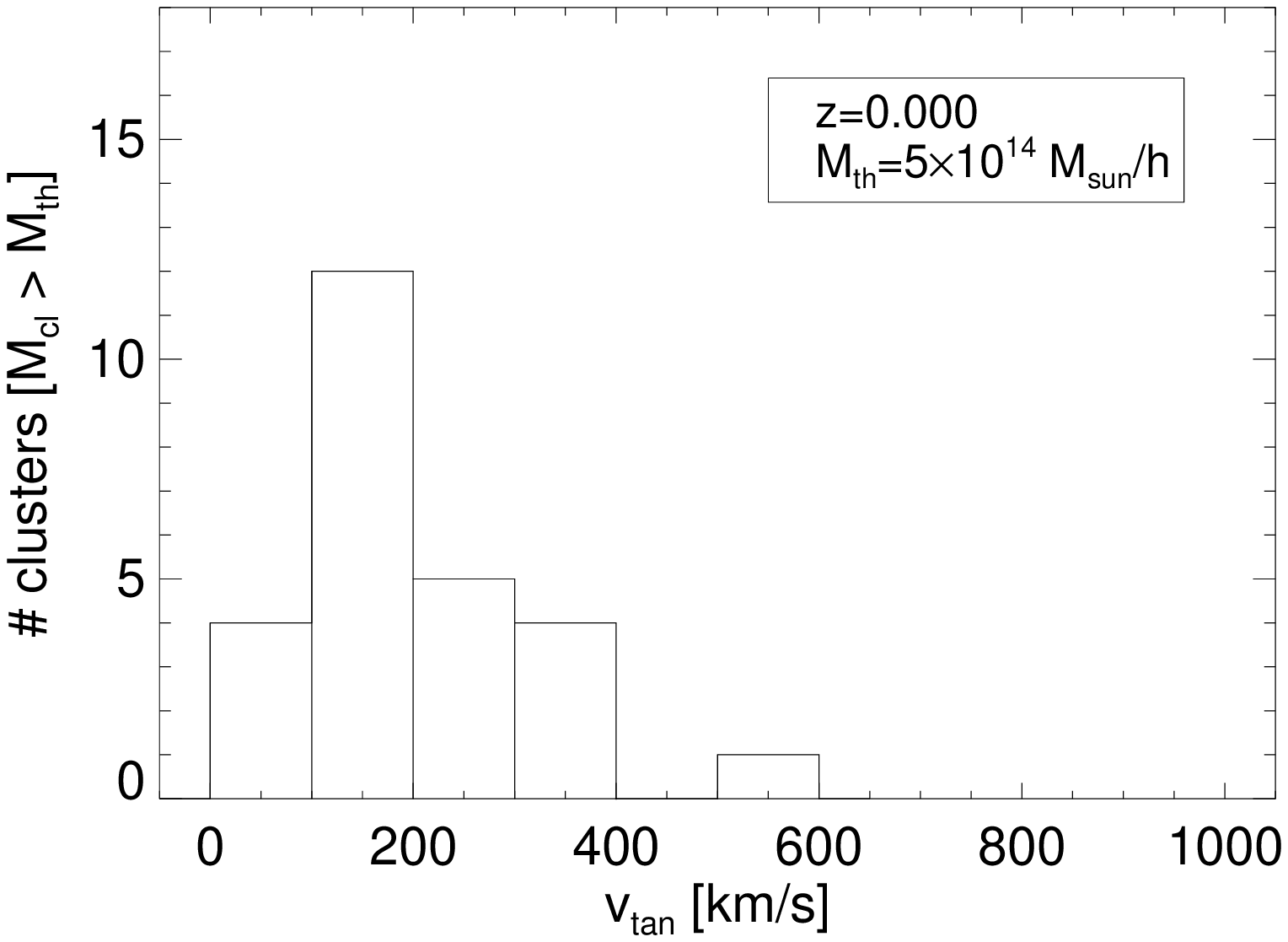}
\includegraphics[scale=0.49]{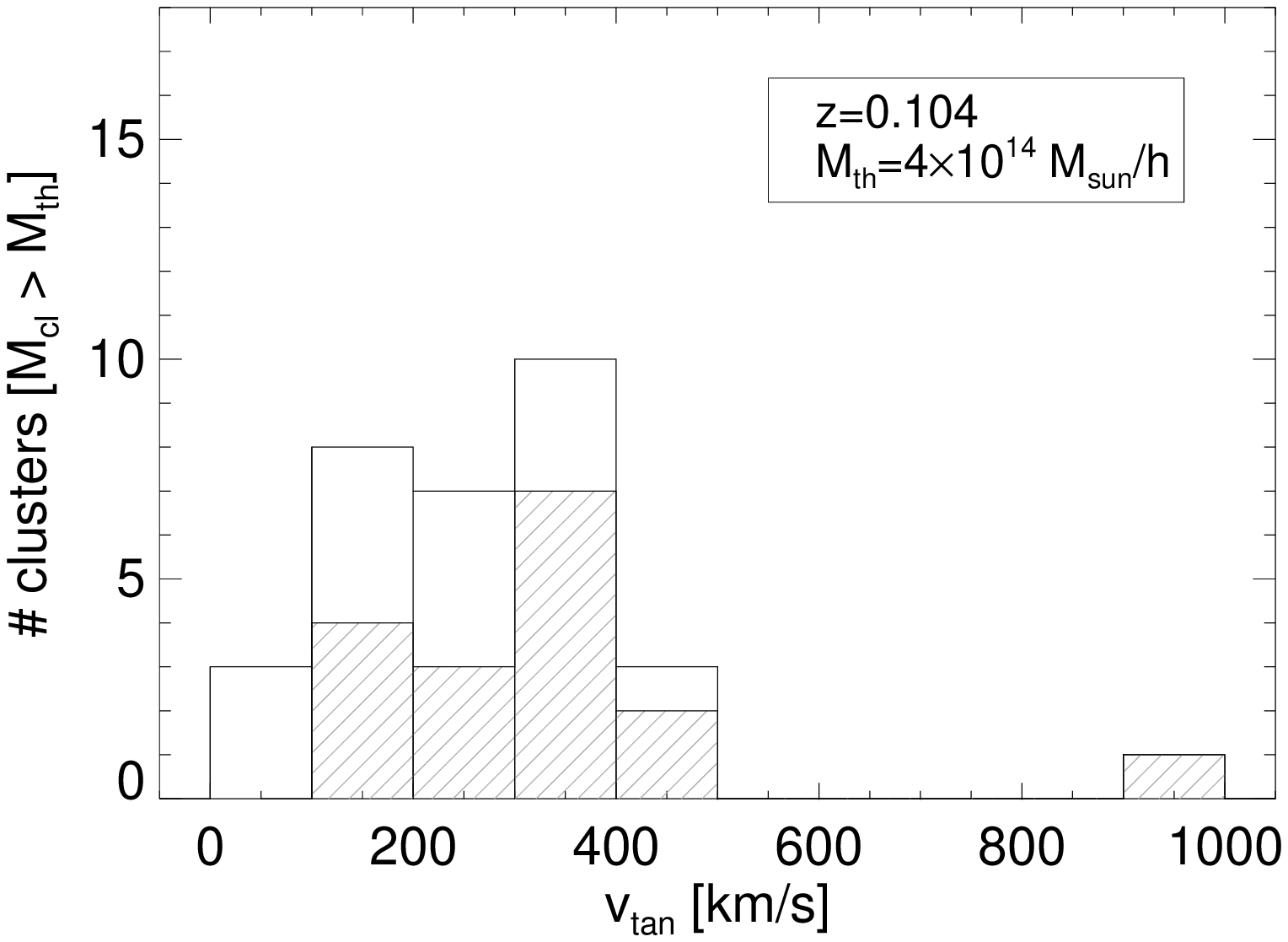}
\includegraphics[scale=0.49]{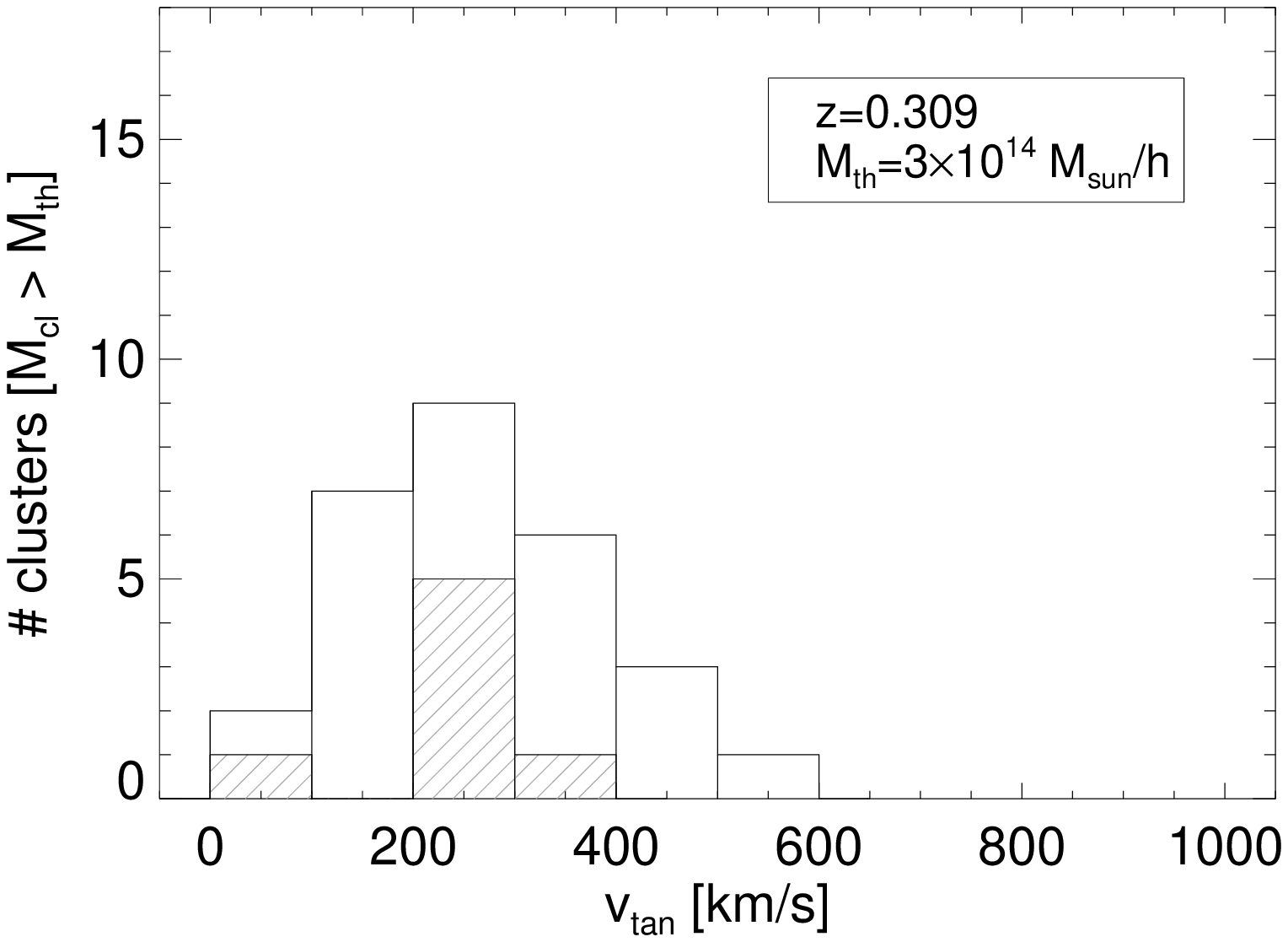}
\includegraphics[scale=0.49]{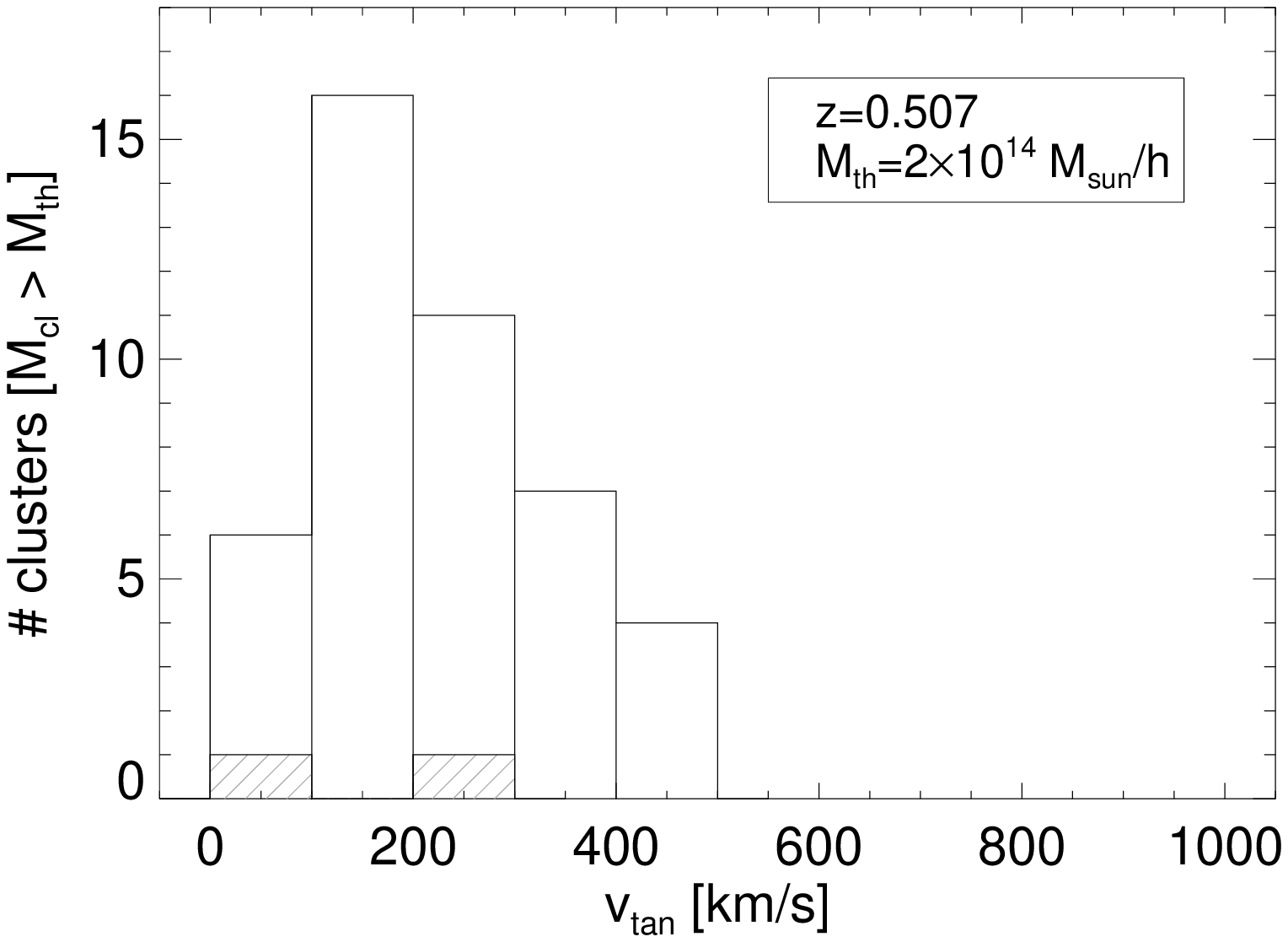}
\caption{Distribution of the rotational velocity for a sample of simulated clusters extracted from the Set 2. The tangential component of the gas velocity, $\vtan,$ is the mean value within $0.1\rfive,$ calculated in the plane perpendicular to the mean gas angular momentum in the same region. Top left to bottom right, the panels refer to a sample of objects from the Set 2 at different redshifts, sampling the range $[0,0.5].$ The sample of objects has been selected according to a threshold in mass, $\mth,$ which is lower at higher redshift in order to keep a fair statistics. While $\mth$ is different at each redshift, the shaded regions represent subsamples of haloes selected to be more massive than $5\times 10^{14}h^{-1}\msun,$ as a comparison to the distribution at $z=0.$}
\label{histo_evol}
\end{figure*}

In \fig\ref{histo_evol} we show the distribution of $\vtan$,
calculated $<0.1\rfive$ in the same way as for g51, for the sample
of cluster--like haloes belonging to the Set 2. The histograms show
how significant rotation is over the range of masses and redshits,
confirming the intermittent nature of the phenomenon. Starting form
the upper--left panel to the bottom--right one, redshift increases
from $0$ to $\sim0.5$ and the sample consists of a number of objects
varying between 26 and 44. The mass threshold chosen, $\mth,$ is
$5\times 10^{14},4\times 10^{14},3\times 10^{14},2\times 10^{14}h^{-1}\msun$ for
the four redshift values considered ($z=0,0.1,0.3,0.5$,
respectively). We notice that in general the velocity distributions
are mainly centered around values of $200-300 \kms,$ with a mild,
though not substantial, shift towards higher values for intermediate
redshift. The shaded histograms on top of the ones plotted for
redshifts $z=0.1,0.3,0.5$ refer to subsamples of haloes with $\mtwom >
5\times 10^{14}h^{-1}\msun$ (the same as the one used at $z=0$). This
comparison is meant to show a clear evidence that also the
higher--mass subsample actually agrees with the general
trend. Although the number of objects in these subsamples decreases
for increasing redshift, this has been done in order to confirm the
idea that indeed the distribution of the rotational velocities is
peaked around quite low values. Except for some outliers, we can
generally exclude any clear monotonic increase of the typical value of
$\vtan$ in the cluster innermost region (always kept to be
$<0.1\rfive$) associated to the assembling of the cluster--size
haloes. In our simulations we cannot find any quiescent build--up
of rotation as a consequence of mass assembly, and the distribution
among the volume--selected sample is not dramatically changing with
redshift.
\section{DISCUSSION AND CONCLUSION} \label{secConclusion}
In this work, we have presented the result of a study over two sets of
hydrodynamical simulations performed with the TreePM/SPH code
GADGET--2. The simulations include radiative cooling, star formation,
and supernova feedback and assume slightly different cosmological
models, a \lcdm one and a WMAP3 one. The main target of this analysis
has been the importance of rotational gas  motions in the central
regions of simulated cluster--like haloes, as it is thought to be a
crucial issue while weighing galaxy clusters and identifying them as
relaxed systems. The objects selected from our samples guarantee a
wide variety of virial masses and dynamical structures, so that a reliable
investigation of this phenomenon is allowed.\\
\indent Our main results can be summarized as follows:
\begin{itemize}
 \item As main conclusion, we notice that the occurrence of rotational
   patterns in the simulated ICM is strictly related to the internal
   dynamics of \textit{gas--rich} substructures in a complicated way, so that
   it is definitely important to take it into account as contribution
   to the pressure support, but it's not directly nor simply connected
   to the global dynamical state of the halo.
 \item In the first part of our analysis we focused on g51, a simulated cluster
   with a very smooth late accretion history, isolated and characterized by few
   substructures in comparison to the other massive objects within Set 1. 
   Also, we compare it with a highly disturbed system (g1). Even in the radiative 
   simulation of this cluster, likely to be considered relaxed in a global sense,
   \textit{no clear rotation} shows up \textit{at low redshift
   because of some minor merging events} occurring \textit{close to
   the innermost region}: the rotation of the core is found to be an
   intermittent phenomenon that can be easily destroyed by the passage
   of gas--rich subhaloes through the equatorial plane. Gas particles
   stripped from the subhalo passing close to the main--halo innermost
   region ($<0.1\rfive$), are likely to get mixed to the gas already
   settled and contribute over few orbits to change the inclination of
   the best equatorial plane, suppressing any pre--existing rotational
   pattern.
 \item The velocity maps plotted in \fig\ref{fig_maps} show several 
   DM--only subhaloes moving close to g51 central core. In our study, they 
   have been found not to disturb in any significant way the ordered rotational
   gas motions created in the innermost region. The central gas sloshing
   is mainly set off by gas--rich subhaloes, especially if they retain their
   gas during the early passages through the core. 
   Interesting work on numerical simulations have been found to be
   relevant for the result presented here, as the study from \cite{ascasibar2006} 
   on the origin of cold fronts and core sloshing in galaxy clusters.
 \item Mass measurements based on HEH are likely to misestimate the
   total mass of galaxy clusters because of contributions by
   non--thermal gas motions that have to be considered. In agreement
   with previous works, we also find that significant rotation of the
   ICM can contribute to the pressure support. 
   While several studies have
   been carried out on turbulent motions in the ICM and on their effect on the cluster
   mass estimates \cite[e.g.][]{rasia04,fang2009,lau2009,zhuravleva2010}, 
   only lately the work by \cite{fang2009}
   and \cite{lau2009} have been addressing the ordered rotational patterns that 
   could establish in the innermost region ICM as the result of the cluster collapse.
   Therefore, a comprehensive analysis of the details of rotation build--up
   and suppression both in single high--resolution case--studies and in larger,
   statistically significant samples is extremely interesting, especially for 
   relaxed objects where this should be more important than turbulence.
   Focusing on rotation specifically, 
   we calculated the corresponding mass term,
   $\mrot,$ for the two clusters g51 and g1. As expected from the
   tangential velocity profiles at redshift $z=0,$ the mass term
   coming from ICM rotational motions contributes more in the case of
   g1 than in g51, providing evidence that rotational support of gas
   in the innermost region is more significant in the former than in
   the latter. While $\mrot$ accounts for few percents at radii close
   to $\rfive$ in both cases, in the central regions up to $\sim17\%$
   of the total true mass in g1 is due to rotational motions of the
   ICM. As regards g51, this contribution is less important, as no
   strong rotation has been found at $z=0,$ but it still reaches a
   value of $\sim10\%$ for the pressure support in the cluster core.
 \item Extending the analysis to a larger sample, we have investigated
   the statistical distribution of rotational velocity over
   dynamically--different clusters, isolated in a limited--volume
   simulated box such that their virial mass ($\mtwom$) is above a
   chosen threshold. At $z=0$ as well as at higher redshift up to
   $\sim0.5,$ a fair sample of cluster--size haloes let us infer that,
   on average, no high--velocity rotational patterns show up in the
   halo cores (i.e. in the region $<0.1\rfive$). Also for the clusters
   of Set 2, we find typical values of $\sim 200-300 \kms$ for the
   rotational velocity in the innermost region.
 \item We do not find any increasing trend of the rotational velocity
   distribution peak with decreasing redshift, that can correspond to
   the smooth mass assembly of the cluster--like halo through collapse. Although
   such trend is generally expected, it must be easily suppressed by
   internal minor events disturbing the halo central region.
\end{itemize}
\indent We conclude that the build--up of rotational patterns in the innermost region of galaxy clusters is mainly 
related to the physical processes included in the \textit{csf} run to describe the intracluster gas. On the contrary, 
numerical effects such as different implementations of artificial viscosity \cite[][]{dolag2005} do not affect in any 
significant way our results (see Appendix~\ref{appendix}, for a detailed discussion).\\
\indent An analogous conclusion can be drawn with respect to the differences between the two samples introduced 
by cosmology and resolution. For both Set 1 and Set 2 the build--up and suppression of rotational patterns in
the halo central part is found to be mainly related to the physics included in the radiative run. 
In fact, comparable subsamples of the two sets in the \textit{csf} simulations show very similar distributions of 
rotational velocities for the ICM component in the halo innermost region, meaning that the shape of the 
distribution is essentially dominated by the physics of the gas.
\indent Usually, relaxed clusters are assumed to have little gas
motions. Therefore they are likely to be the best candidates for the
validity of the HEH, on which mass estimations are
based. Nevertheless, rotational motions should establish
preferentially in relaxed clusters with respect to disturbed systems
as a consequence of the assembling process, potentially representing a
danger for relaxed cluster masses. Here, however, we find that the 
processes described in the paper save the
reliability of the HEH--based mass determinations in most of the
cases. In fact, rotational motions are not significant enough to
compromise dramatically mass determinations with the exception of few
outliers. In our simulation, the identification of relaxed or
non--relaxed clusters according to the presence of gas rotation in the
central region is not straightforward, since it has been shown to
appear and disappear periodically. Its contribution has to be
considered whenever is present, but it is not directly related to the
global state of the simulated halo. Also, it is likely to be strongly
influenced by the overcooling problem affecting hydrodynamical
simulations, which has the effect to enhance the process of building up 
rotational patterns in the ICM in the innermost regions of
simulated clusters.\\
\indent
Although various theoretical and numerical studies in
addition to the present work have been investigating the existence of gas bulk, non--thermal
motions and the possible ways to detect them in galaxy clusters 
\cite[e.g.][]{fang2009,lau2009,zhuravleva2010}, little is
known from observations. In a recent study, \cite{lagana2009} have
made use of assumptions from theoretical models and numerical
simulations about cosmic rays, turbulence and magnetic pressure to
consider these non--thermal contributions to the total mass
measurement for five Abell clusters. From a pure observational point
of view, previous work has been able to confirm only indirect
indications of bulk gas motions associated to merging events in galaxy
clusters \cite[see][for a review]{markev2007} or evidences for
turbulent gas motions, like the ones found in the Coma cluster in
\cite{schuecker2004} or those inferred, on the scale of smaller--mass
systems, from the effects of resonant scattering in the X-ray emitting
gaseous haloes of large elliptical galaxies
\cite[][]{werner2009}. Although not possible so far, the most direct
way to measure gas motions in galaxy clusters would be via the
broadening of the line profile of heavy ions (like the iron line at
$\sim6.7\kev$ in X--rays) for which the expected linewidth due to
impact of gas motion is much larger than the width due to pure thermal
broadening. The possibility to use the shape of the emission lines as
a source of information on the ICM velocity field as been discussed in
detail in \cite{inogamov2003} and \cite{sunyaev2003}, and lately in
\cite{rebusco2008}. Though, the investigation of the imprint of ICM
motions on the iron line profile requires high--resolution
spectroscopy, which will become possible in the near future with the
next--generation X--ray instruments such as ASTRO--H and IXO. This will
allow us to directly detect non--thermal contributions to the cluster
pressure support, such as rotational patterns in the ICM, and enable
us to take this correctly into account as contribution to the total
mass estimate. Ultimately, this is likely to be an important issue to
handle in order to better understand deviations from the HEH, on which
scaling laws are usually based.

\section*{Acknowledgments} 
The simulations have been performed at the “Leibniz-Rechenzentrum” with CPU time 
assigned to the Project “h0073.” KD acknowledges the support of the DFG 
Priority Program 1177. HB acknowledges the support by the Excellence Cluster 153 
supported by the German Federal Government. We want to thank Eugene Churazov and 
Irina Zhuravleva for useful discussions that helped to improve the manuscript. 
VB gratefully acknowledges Lodovico Coccato for help with IRAF. 
We wish also to thank the anonymous referee for helpful comments that improved the 
presentation of our results.

\appendix
\section[]{Effects of artificial viscosity}\label{appendix}
\begin{figure}
	\includegraphics[scale=0.49]{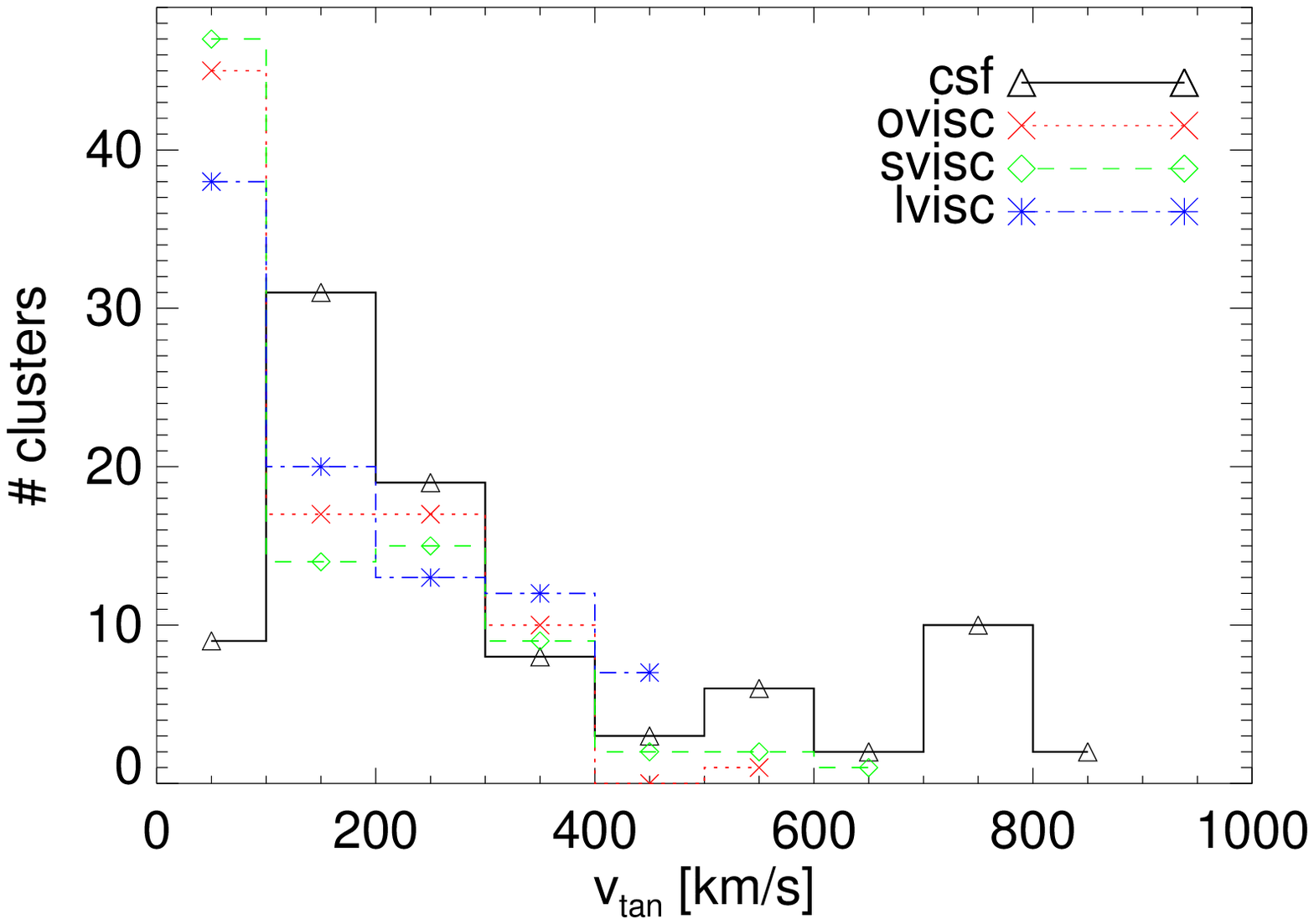}
	\caption{Distribution of rotational velocity (calculated in the region $<0.1\rfive$) for the Set 1 in the redshift range $[0-0.5]$. The different runs are plotted: \textit{csf} (\textit{solid, black line} with \textit{triangles}), \textit{ovisc} (\textit{dotted, red} with \textit{crosses}), \textit{svisc} (\textit{green, dashed} with \textit{diamonds}), \textit{lvisc} (\textit{blue, dot-dashed} with \textit{asterisks}).}
	\label{histohutt}
\end{figure}

The runs studied in the present work are
the \textit{csf} simulation, including radiative cooling, star formation and supernova feedback,
and the non--radiative simulation (labelled as \textit{ovisc}), where the original parametrization
of artificial viscosity by \cite{monaghan83} is used. We comment here on the effects of the artificial 
viscosity on our study by considering two additional non-radiative runs of our simulations, carried out with 
alternative implementations of the artificial viscosity scheme. In particular, we label as \textit{svisc}
the non-radiative run with slightly less numerical viscosity based on the signal velocity approach of
\cite{monaghan97}, and as \textit{lvisc} the modified artificial viscosity scheme where each particle
evolves with its own time--dependent viscosity parameter (as originally suggested in \cite{morris97}).
For a detailed description of these non-radiative runs we refer the reader to \cite{dolag2005}.\\
\indent In \fig\ref{histohutt} the distribution of the value of $\vtan$ for the Set 1 is presented for
the different runs. With the solid black line we refer to the \textit{csf} simulation, while the other
histograms represent the non-radiative runs: \textit{ovisc} (dotted red line), \textit{svisc} (green dashed) 
and \textit{lvisc} (blue, dot--dashed). As one can see from the Figure, the \textit{csf} simulation shows a
different distribution of rotational velocity in the innermost regions of the cluster--like haloes
with respect to the non--radiative runs, in which the difference in the implementation for the artificial
viscosity does not produce significant differences in the three distributions. 
In \fig\ref{histohutt} is evident that the difference between the radiative and non--radiative runs is much larger 
than the difference among the non--radiative runs themselves.
All the non--radiative runs similarly
show that the largest fraction of clusters have very low values of $\vtan,$ and differ mostly in the
lack of a high-velocity end of the distribution from the \textit{csf} case. 
The cooling of the core allowed by the physics included in the \textit{csf} run is plausibly 
responsible for the presence of a significant fraction of
clusters with high velocity values, which do not exist in the non--radiative runs.\\
\indent From \fig\ref{histohutt} we can confirm that, accordingly to what is expected, 
the overcooling problem coming along with numerical simulations of galaxy clusters leads to a more 
significant build--up of rotation in the core, reflected in an overall shift towards higher values 
of the distribution of $\vtan.$\\
\indent For the purpose of our study, we can conclude that the main effects in the establishment of
rotational patterns in the central region of simulated clusters are introduced by the physical
processes describing the gaseous component, included in the \textit{csf} simulation. 
The amount of turbulence, referring in particular to small scale chaotic motions, was found to strongly differ
among the non--radiative runs investigated here \cite[e.g.][]{dolag2005}. 
However, the effect of artificial viscosity on rotation does not depend 
strongly on the specific numerical scheme implemented,
and we safely confirm our conclusions about the more significant effect that minor events 
occurring close to the cluster central region have on the survival of gas rotational motions.\\
\section[]{Ellipticity of ICM}\label{appendix2}
\begin{figure*}
	\includegraphics[scale=0.49]{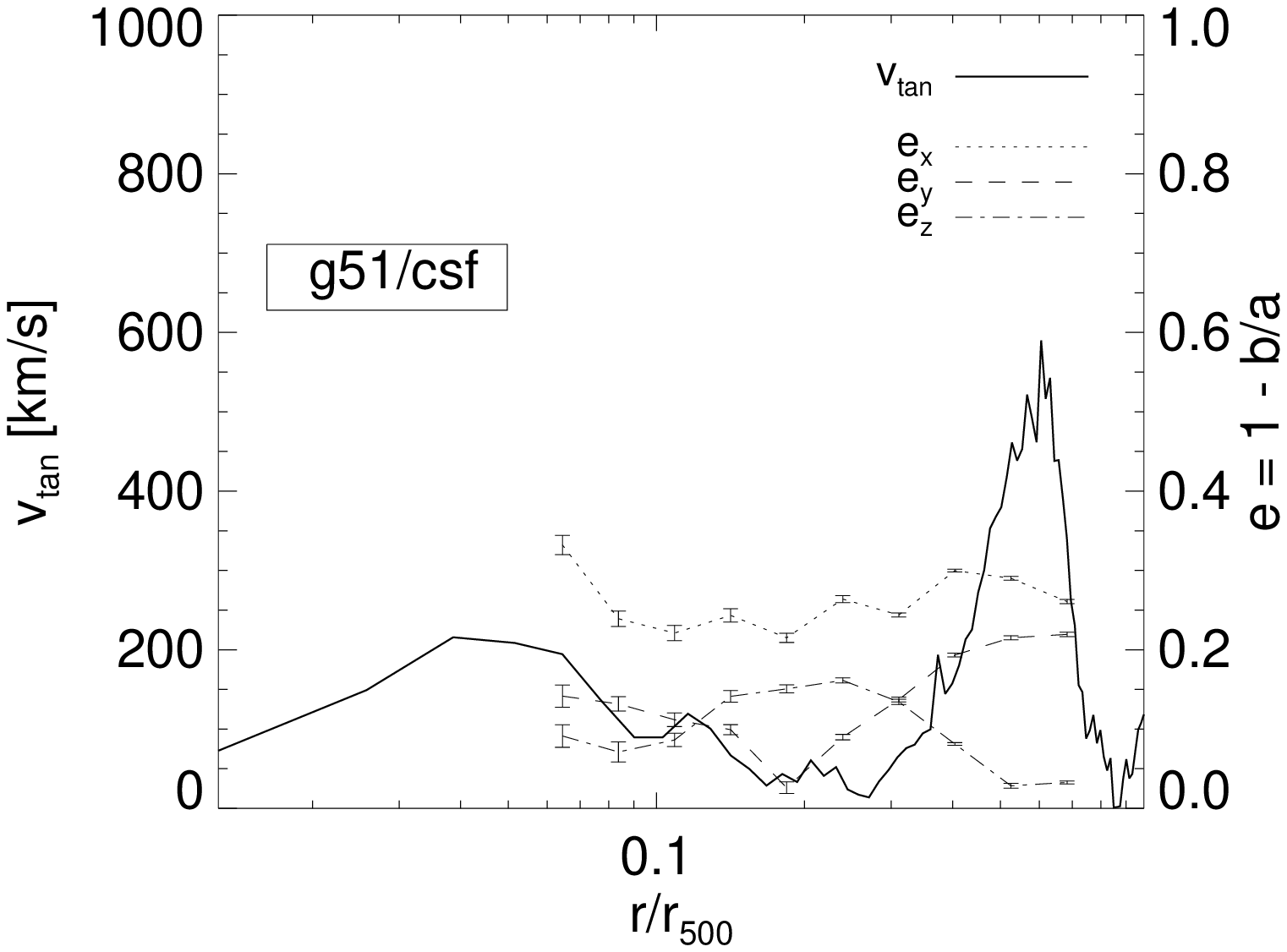}
	\includegraphics[scale=0.49]{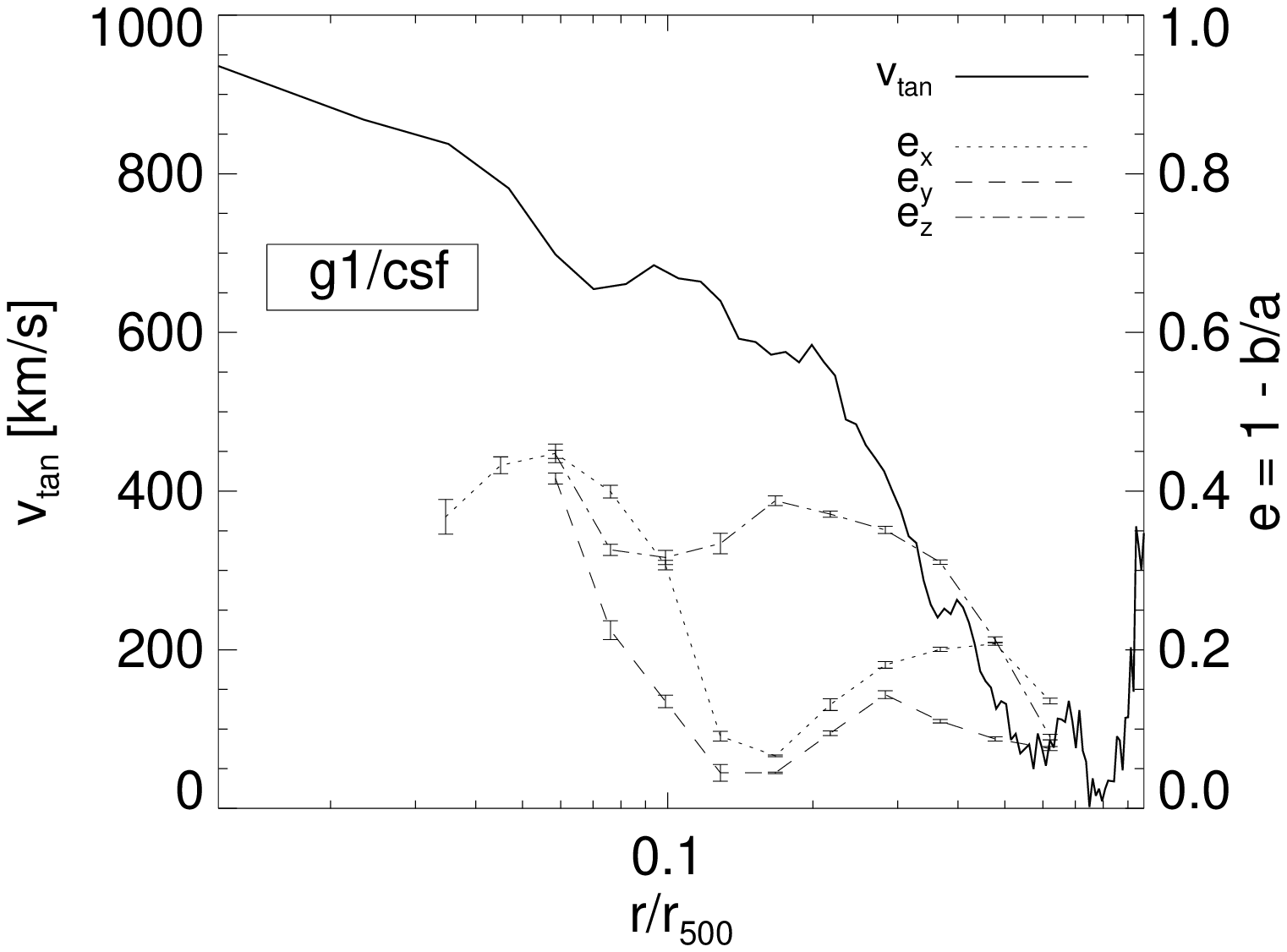}
	\includegraphics[scale=0.49]{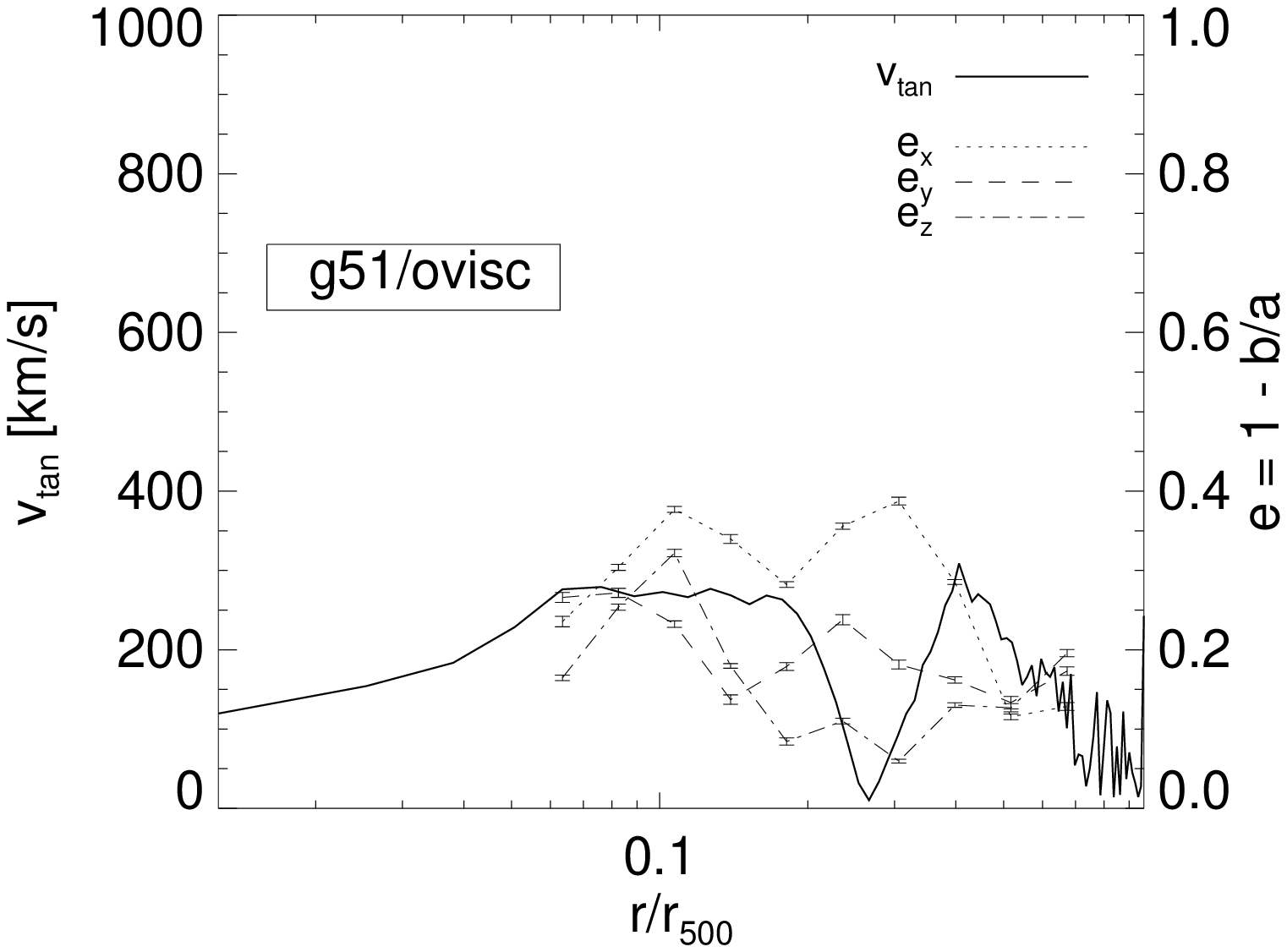}
	\includegraphics[scale=0.49]{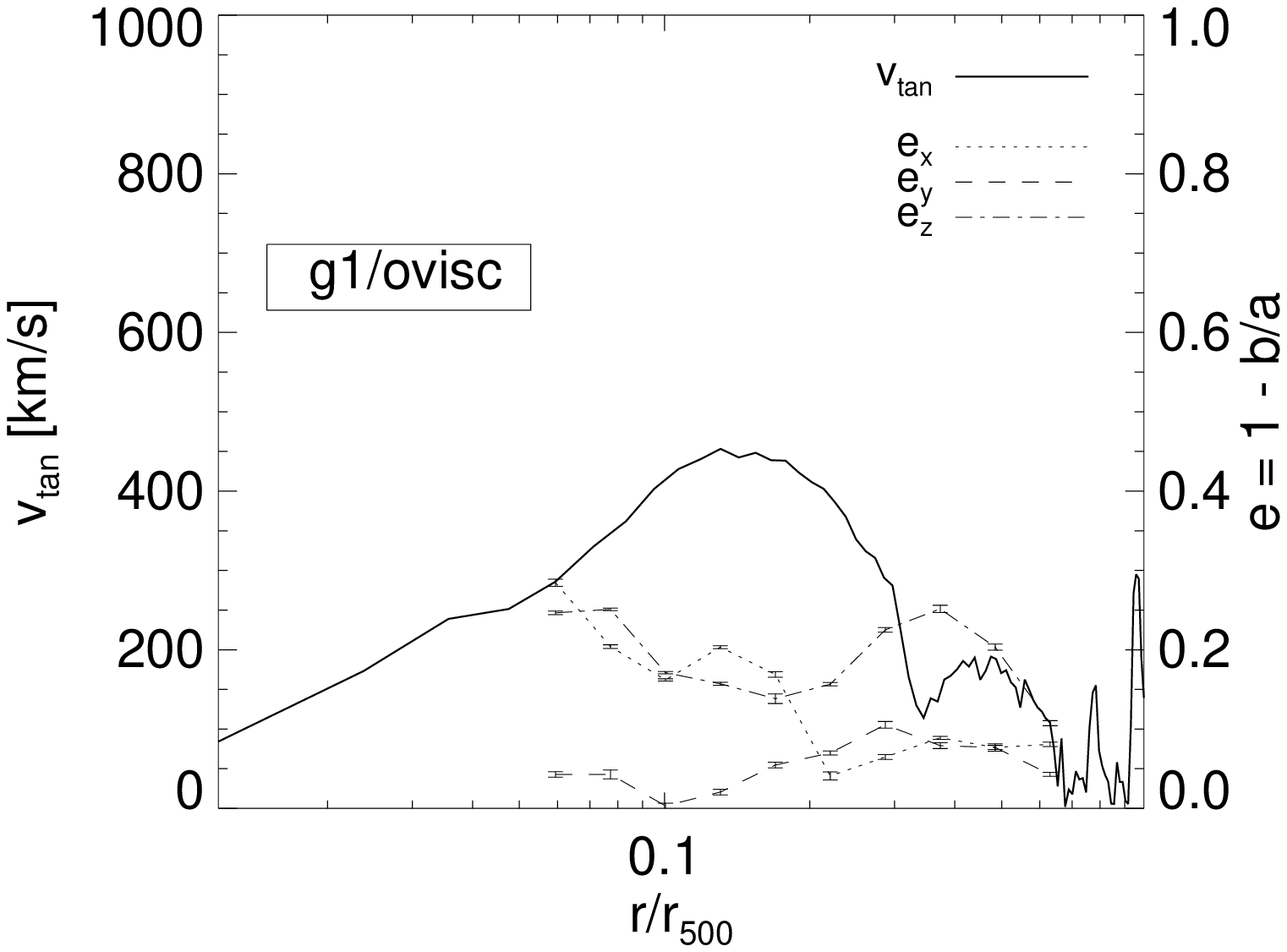}
	\caption{Radial profiles of ellipticity (\textit{left y--axis; dotted, dashed and dott--dashed curves}) for the three projection axis compared with the radial profile of the gas rotational velocity (\textit{right y--axis; solid curve}). The panels refer to the \textit{csf} and the \textit{ovisc} simulations (\textit{upper} and \textit{lower} panels, respectively) of g51 (\textit{left}) and g1 (\textit{right}) at $z=0$.}
	\label{ellip}
\end{figure*}

As interestingly suggested in a recent work by \cite{lau2010}, certain observable features of the cluster 
X--ray emission, like a flattening of cluster shapes, could unveil the presence of rotationally supported gas, especially in the innermost region. 
Referring to the X--ray surface brightness maps in \fig\ref{hutt_maps}, we compare here the ellipticity 
profiles extracted from the maps in the three projections to the rotational velocity profile, 
for the two cases of study extracted from Set 1 (namely, g51 and g1) at redshift $z=0$. 
The ellipticity was calculated by means of the IRAF task \texttt{ellipse} 
and the reported error bars are those obtained by the fitting method. 
In \fig\ref{ellip} the $\vtan$ profile 
is marked by the solid curve for each cluster, while the dotted, dashed and dot--dashed lines refer to the 
ellipticities from the three projected maps. Let us note that the ellipticity profiles do not extend to the 
very central part, since the complex structure in the innermost region does not allow
for a simple determination of ellipticities. 
We can conclude that only a mild relation between ellipticity 
and rotation is found in these clusters and the difference in the trends shown by the rotational velocity 
profiles is stronger than the difference among the ellipticity profiles of the two clusters. 
Nevertheless, comparing the \textit{csf} run (upper panels) to the \textit{ovisc} run (lower panels) we can confirm that the cluster shape 
is actually marked by the physics included in the radiative run, especially in the g1 case, where some 
rotation establishes in the \textit{csf} simulation.
In conclusion, gas cooling can affect the gas shape by flattening the X--ray isophotes but we do not find 
a strong significant evidence for that in our case--study clusters, in agreement with the milder rotation
found and the nature of such rotational patterns (like in the g1 example), which are likely to be temporary effects
driven by major merging events.

\bsp

\label{lastpage}

\end{document}